\documentclass{ieeeaccess}
\usepackage{cite}
\usepackage{amsmath,amssymb,amsfonts}
\usepackage{graphicx}
\usepackage{textcomp}
\usepackage{bm}
\usepackage{cite}
\usepackage{tikz}
\usepackage{pgfplots}
\pgfplotsset{compat=1.18}
\usepackage{booktabs}
\usepackage{multirow}
\usepackage{algorithm}
\usepackage{algpseudocode}
\usepackage{hyperref}
\usepackage{tabularx}
\usepackage{subcaption}
\usepackage[table]{xcolor}

\NewSpotColorSpace{PANTONE}
\AddSpotColor{PANTONE} {PANTONE3015C} {PANTONE\SpotSpace 3015\SpotSpace C} {1 0.3 0 0.2}
\SetPageColorSpace{PANTONE}

\makeatletter
\AtBeginDocument{\DeclareMathVersion{bold}
\SetSymbolFont{operators}{bold}{T1}{times}{b}{n}
\SetSymbolFont{NewLetters}{bold}{T1}{times}{b}{it}
\SetMathAlphabet{\mathrm}{bold}{T1}{times}{b}{n}
\SetMathAlphabet{\mathit}{bold}{T1}{times}{b}{it}
\SetMathAlphabet{\mathbf}{bold}{T1}{times}{b}{n}
\SetMathAlphabet{\mathtt}{bold}{OT1}{pcr}{b}{n}
\SetSymbolFont{symbols}{bold}{OMS}{cmsy}{b}{n}

\def\thevol{xx}
\def\theyear{2026}

\renewcommand\boldmath{\@nomath\boldmath\mathversion{bold}}}
\makeatother

\def\BibTeX{{\rm B\kern-.05em{\sc i\kern-.025em b}\kern-.08em
    T\kern-.1667em\lower.7ex\hbox{E}\kern-.125emX}}

\begin{document}
\history{Date of publication xxxx 00, 0000, date of current version xxxx 00, 0000.}
\doi{10.1109/ACCESS.20XX.XXXXXX}

\title{Bayesian and Multi-Objective Decision Support for Incident Mitigation in Cyber--Physical Systems}
\author{\uppercase{Shaofei Huang}\authorrefmark{1},
\uppercase{Christopher M. Poskitt\authorrefmark{2}, and Lwin Khin Shar}\authorrefmark{3}}
\address[1]{Singapore Management University, Singapore (email: sf.huang.2023@smu.edu.sg)}
\address[2]{Singapore Management University, Singapore (email: cposkitt@smu.edu.sg)}
\address[3]{Singapore Management University, Singapore (email: lkshar@smu.edu.sg)}

\markboth
{Huang \headeretal: Bayesian and Multi-Objective Decision Support for Incident Mitigation in Cyber-Physical Systems}
{Huang \headeretal: Bayesian and Multi-Objective Decision Support for Incident Mitigation in Cyber-Physical Systems}

\corresp{Corresponding author: Shaofei Huang (email: sf.huang.2023@smu.edu.sg).}

\begin{abstract}

Cyber-physical systems increasingly rely on interconnected physical and digital systems whose security incidents can escalate rapidly into safety and operational failures. Existing decision-support approaches struggle to support incident response because they rely on static assumptions, incomplete vulnerability data, and single-objective risk models that do not adequately capture trade-offs between attack success likelihood, impact severity, and system availability. This paper proposes an adaptive decision-support framework for incident mitigation in cyber-physical systems that integrates hierarchical Bayesian Network modelling, confidence-calibrated exposure estimation, and multi-objective optimisation into a unified, adaptive pipeline. The framework constructs probabilistic models from system architecture and vulnerability data, incorporating complementary vulnerability scores under epistemic uncertainty as conservative, uncertainty-aware reporting metrics for supporting downstream risk assessment. Mitigation strategies are explored as countermeasure portfolios and refined using multi-objective optimisation to identify Pareto-optimal trade-offs suitable for incident response scenarios. Frequency-based heuristics are applied to prioritise mitigation actions across optimisation runs. The framework is evaluated on three representative cyber-physical attack scenarios, demonstrating its ability to adapt to evolving threats and provide actionable decision support under operational constraints, with the aim of enhancing the resilience of cyber-physical systems.

\end{abstract}


\titlepgskip=-21pt

\maketitle

\section{Introduction}
\label{sec:introduction}

Cyber-Physical Systems (CPSs) form the backbone of critical infrastructure, integrating digital and physical components to deliver the essential services that underpin modern society. While traditionally confined to centralised Industrial Control Systems (ICSs), CPSs are now deployed at scale across decentralised infrastructure such as railway signalling networks and renewable energy ecosystems. This expansion has significantly broadened the attack surface for both state-sponsored and opportunistic actors, as digital vulnerabilities now translate directly into physical hazards and operational disruptions that threaten national security and public safety.

A pivotal example of this shifting threat landscape was highlighted in 2025 by Dashevskyi et al. in their SUN:DOWN research report~\cite{Dashevskyi2025}, which disclosed 46 new vulnerabilities across leading solar photovoltaic (PV) inverter vendors. These flaws, ranging from insecure cloud APIs to remote code execution, could be chained to orchestrate a coordinated load-changing attack, manipulating inverter output to destabilise grid frequency and potentially triggering large-scale load shedding or emergency equipment shutdowns, underscoring the systemic risk unsecured CPS components pose to infrastructural resilience.

Given the critical nature of these processes, \emph{adaptive} decision support is essential to ensure robust security and incident mitigation, while simultaneously minimising impacts on safety and preserving continuous system availability. In this context, ``adaptive'' refers to the capacity to dynamically adjust mitigation strategies based on evolving threat conditions, system states, and resource availability. Unlike traditional IT environments, where remediation typically involves rapid software patching, CPS security decisions must balance cybersecurity objectives against physical process constraints~\cite{Taylor2017}. In legacy CPSs, applying an unvalidated patch entails substantial operational risk. Updates may inadvertently result in latent system failures or performance degradation, necessitating countermeasures that prioritise the resilience of critical infrastructure systems~\cite{Rehak2019}.

Despite advancements in model-driven decision support, existing frameworks often operate under static assumptions and lack the adaptability required during an unfolding incident~\cite{Li2018,Zaman2022}. Most frameworks cannot update model parameters such as vulnerability exposure probabilities or asset interdependencies dynamically. A representative scenario is the detection of a compromised communication dongle mid-response in a power grid. The inability to dynamically revise model attributes constrains operational decision-making and reduces response efficacy in operationally critical contexts.

\begin{figure}[t]
    \centering
    \includegraphics[width=\linewidth]{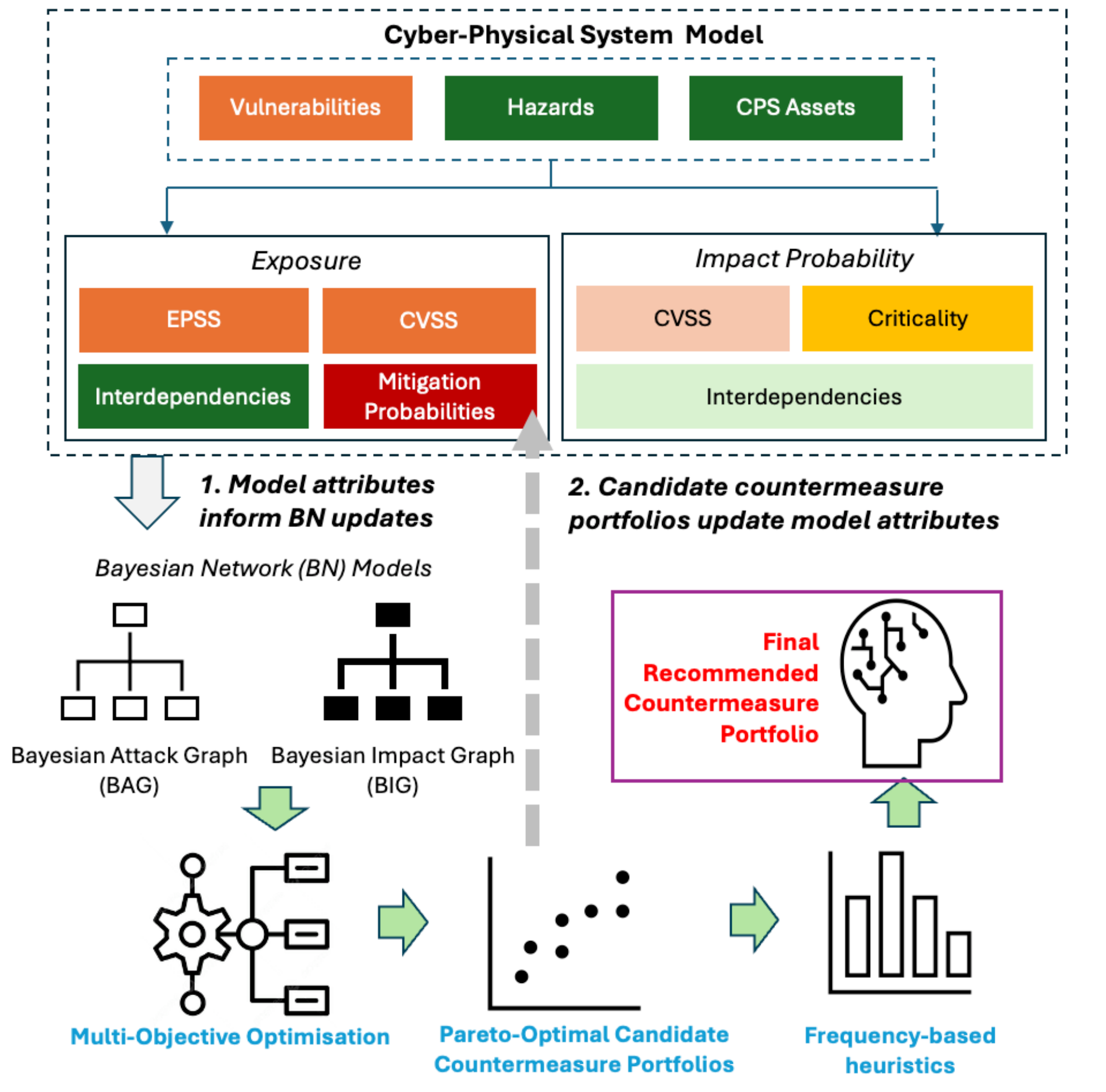}
    \caption{Architecture of our proposed CPS security framework. The diagram shows an iterative process integrating a CPS model (vulnerabilities, hazards, assets, exposure/impact probability attributes) with BN graph construction (BAG and BIG). Multi-objective optimisation generates Pareto-optimal countermeasure portfolios, updating model attributes and propagating exposure/impact probabilities. Frequency heuristic analysis selects the optimal portfolio.}
    \label{fig:framework_architecture}
\end{figure}

To address these limitations, we propose an adaptive decision-support framework for incident mitigation in CPSs, grounded in applied probabilistic reasoning and designed to meet the operational realities of CPS protection. We posit that effective response strategies must be guided by data-driven risk metrics while remaining bounded by stringent operational constraints. As illustrated in Figure~\ref{fig:framework_architecture}, our proposed framework integrates Bayesian Network (BN) graphs for CPS representation and risk assessment~\cite{Poolsappasit2011}, with multi-objective optimisation frameworks to elicit Pareto-optimal countermeasures for incident response \cite{Li2018,Zebrowski2022}.

\textit{Novelty}. The key novelty of this work lies in the integration of confidence-calibrated exposure estimation, BN-based risk propagation, and multi-objective optimisation into a unified, adaptive decision-support pipeline, further augmented by frequency-based heuristics for countermeasure prioritisation. This pipeline renders the Bayesian structures established by Poolsappasit et al.~\cite{Poolsappasit2011}, alongside the multi-objective optimisation frameworks proposed by Li et al.~\cite{Li2018} and {\.Z}ebrowski et al.~\cite{Zebrowski2022}, actionable under real-world incident response constraints. Whilst these prior works largely assume well-specified probabilities and static or offline decision contexts, our approach explicitly incorporates \emph{epistemic uncertainty} and \emph{model confidence} into the exposure computation itself, bounding the influence of unreliable proxy scores and producing uncertainty-aware reporting metrics that support downstream risk assessment.

To the authors' knowledge, while commercial platforms such as Dragos and Claroty use machine-learning-based exploitability scoring and expert-driven risk models, they do not explicitly formalise epistemic uncertainty within their exposure metrics~\cite{Claroty2026, Dragos2024}. In contrast, our proposed confidence-calibrated metrics provide practitioners with a supporting component to inform risk assessments, whilst its formal integration into the decision-support pipeline remains a direction for future work.

\textit{Contributions}. The main contributions of our paper are summarised below:

\begin{itemize}
    \item We present a BN-based modelling approach designed explicitly for adaptive incident response in CPSs, rather than offline risk analysis. The model embeds system and vulnerability metadata in a Domain-Specific Language (DSL)~\cite{DSL2010}, enabling rapid updates to structure and parameters as incidents unfold.

    \item We develop a hybrid exposure probability estimation framework that integrates complementary vulnerability scores through a Bayesian confidence calibration technique. The framework bounds the influence of unreliable proxy scores and produces uncertainty-aware reporting metrics that support downstream risk assessment.

    \item We introduce a multi-objective decision-support methodology for incident mitigation in cyber-physical systems that balances security, safety, and operational resilience. By treating mitigation probabilities as dynamic decision variables, the framework identifies Pareto-optimal \textit{countermeasure portfolios} tailored to CPS constraints, and uses frequency-based heuristics to prioritise mitigation strategies under operational conditions.

    \item We demonstrate the applicability of our proposed framework through three representative case studies across diverse CPS domains: the 2015 Ukrainian power grid (BlackEnergy) attack, a solar PV inverter system, and a railway CBTC (Communications‑Based Train Control) network.

    \item The models and source code developed in this research, along with the reference data and formulae used in the framework, are publicly available via the project’s GitHub repository~\cite{Huang_GitHub}. This supports transparency, facilitates reproducibility and enables further development by both the research and practitioner communities within the CPS protection domain.

\end{itemize}

The remainder of this paper is structured as follows. Section~\ref{sec:background} provides background on the key techniques adopted in this work. Section~\ref{sec:methodology} explains the proposed methodology. Section~\ref{sec:evaluation} presents applications of the method to CPS attack scenarios, accompanied by an analysis of experimental results. Section~\ref{sec:discussion} provides an in-depth, technical discussion of the findings and outlines directions for future research. Section~\ref{sec:related_work} reviews related research and identifies limitations in existing approaches. Section~\ref{sec:conclusion} summarises the key contributions of the study.

\section{Background}
\label{sec:background}

This section outlines the foundational techniques that underpin our proposed framework.

\paragraph{Bayesian Network Graphs}

BN graphs are probabilistic graphical models that represent dependencies among variables using directed acyclic graphs, as illustrated in Figure~\ref{fig:case_SolarPV_BN}. In cybersecurity, BN graphs are widely used to model attack propagation and risk inference~\cite{Kim2022}, and also to support decision-making~\cite{Javornik2022}. Specific to CPS, the works by Bhosale et al.~\cite{Bhosale2023, Bhosale2024} discuss the adaptation of BN graphs to model probabilistic interdependencies between node types associated with CPS, namely, vulnerability nodes, which represent exploitable weaknesses in systems; asset nodes, which denote critical components within the CPS; and hazard nodes, which correspond to severe adverse outcomes including system failures or performance degradation.

Each node is associated with \textit{exposure} and \textit{impact} attributes that quantify the likelihoods of exploitation, failure dependencies, and cascading risks. These attributes shape the propagation of vulnerabilities and attack vectors, influencing the probabilities of attack success and of severe impact to the CPS. Our work builds on these foundations by integrating such BN-based exposure and impact modelling into an adaptive decision-support pipeline for incident mitigation, and by coupling the BN graph with optimisation and heuristic prioritisation to drive operational countermeasure selection rather than using the BN graph solely for static risk analysis.

\paragraph{Domain Specific Languages}

DSLs are concise, processable languages for describing a specific concern when building a system in a specific domain, where the abstractions and notations used are tailored to the stakeholders who specify that particular concern \cite{DSL2010}. In this work, we employ AutomationML~\cite{AutomationML2008} (an XML-based DSL for representing engineering models) as a means of encoding and representing BN graphs for automated analysis. Using a DSL such as AutomationML allows CPS practitioners and experts to formally specify CPSs in a BN graph structure (nodes, edges, and conditional probability tables) along with vulnerabilities, assets, and countermeasures in a machine-readable format.

\paragraph{Vulnerability Scoring Systems}

The Common Vulnerability Scoring System (CVSS) provides a standardised framework for assessing the technical severity of vulnerabilities, with version 3.1 defining metrics to quantify exploitability and impact characteristics~\cite{CVSS31,CVSS2007}. However, as a deterministic scoring system, CVSS often fails to account for the inherent uncertainty and temporal dynamics of real-world attacks. The Exploit Prediction Scoring System (EPSS) addresses this limitation by providing probabilistic estimates of exploitation likelihood over the subsequent 30-day window, grounded in empirical indicators such as exploit availability and active exploitation trends~\cite{EPSS2021}.

To achieve a more granular assessment, our framework further integrates the \textit{Known Exploited Vulnerabilities} (KEV) catalogue~\cite{KEV2026}, which identifies vulnerabilities confirmed to have been exploited in real-world scenarios, alongside \textit{Likely Exploited Vulnerabilities} (LEV)~\cite{Mell2025} data, which provides daily probability estimates of observed exploitation activities. Our approach captures a broader spectrum of the threat landscape by incorporating both authoritative records and statistical exploitation estimates.

While these scoring frameworks are well-established within the cybersecurity community, our framework distinguishes itself by integrating them as complementary scoring schemes within a Bayesian calibration mechanism, yielding uncertainty-aware metrics that explicitly account for incomplete or heterogeneous vulnerability metadata and support our decision-support process.

\paragraph{Multi-Objective Optimisation}

CPS decision makers are often constrained by fixed budgets that fall short of the minimum cost required for comprehensive system hardening. The challenge lies in selecting a subset of security measures that remain within budget while minimising residual risk from unaddressed vulnerabilities~\cite{Dewri2007}. Multi-objective optimisation addresses this challenge by identifying solutions that balance competing objectives, such as cost, system availability, and risk reduction. Prior research has explored Bayesian optimisation and evolutionary strategies to derive optimal security configurations under uncertainty and resource constraints~\cite{Li2018,Dewri2007}. In particular, the work by {\.Z}ebrowski et al.~\cite{Zebrowski2022} formulates the optimisation task as a Pareto-efficient portfolio selection problem, leveraging a Bayesian Network to identify security measures that simultaneously minimise various types of expected cyberattack impacts, while satisfying budgetary and other constraints. Our work extends this approach by treating mitigation probabilities as parameters that shape the decision variables within the optimisation space, repeatedly updating BN graph parameters in response to candidate portfolios, and applying frequency-based heuristic analysis across multiple optimisation runs to prioritise countermeasures under operationally-constrained incident response conditions.

\section{Methodology}
\label{sec:methodology}

This section describes the methodology of our proposed framework, illustrated with a running example of a solar PV inverter attack, which is based on the security reports by Dashevskyi et al.~\cite{Dashevskyi2025} and Dabrowski et al.~\cite{Dabrowski2017}. In particular, we refer to their identified CVE (Common Vulnerabilities and Exposures) entries that can be chained to form hypothetical attack paths targeting solar PV inverter systems. While Dabrowski et al.~originally proposed destabilising the power grid via coordinated load modulation from a botnet of consumer devices, our scenario explores a complementary threat vector focused on the supply side. Here, the adversary leverages mobile or web applications to obtain sensitive information, which is then used to compromise a message broker and subsequently a controller linked to both the message broker and a solar PV inverter. Once initial access is established, the attacker hijacks additional inverters and synchronises their configuration changes to create abrupt fluctuations in grid input. By overwhelming the grid’s balancing mechanisms, this coordinated attack can degrade grid stability or result in power outages, thereby achieving the attacker’s ultimate objective.

\emph{Overview.} Our framework (Figure~\ref{fig:framework_architecture}) 
begins with the construction of a BN graph that defines a shared structure for attack and impact graphs, informed by CPS attributes and the attack scenario. Randomised mitigation probability values are introduced as initial inputs into a multi-objective optimisation process, which generates candidate countermeasure portfolios aimed at maximising system availability while minimising the likelihood of attack success and severe impact occurrence. These portfolios are subsequently reintegrated into the BN graph as updated mitigation probabilities, and the optimisation cycle is repeated across multiple trials within a single run. The top-performing portfolios from these iterative runs are then analysed using frequency-based heuristics to identify vulnerability nodes within the attack path that consistently exhibit high mitigation effectiveness. This insight serves to continually guide and inform final prioritisation decisions during incident response. The high-level steps of our proposed framework are outlined in Algorithm~\ref{alg:framework_overview}. Each step is elaborated in detail in Sections~\ref{subsec:method_bn_construction} to \ref{subsec:method_freq_based_analysis}.

\begin{algorithm}
\caption{Methodology of Proposed CPS Decision Support Framework}
\begin{algorithmic}[1]
    \footnotesize
    \State \textbf{Build} BN graph integrating CPS metadata and attack scenario context, encoded in AutomationML (§~\ref{subsec:method_bn_construction})
    
    \State \textbf{Compute} exposure probabilities for nodes within the BN graph, representing likelihoods of attack success (§~\ref{subsec:method_exposure_calculation})
    
    \State \textbf{Calculate} propagated impact probabilities resulting from exploitation or failure of BN graph nodes (§~\ref{subsec:method_impact_calculation})
    
    \State \textbf{Calculate} posterior likelihoods of successful attack and severe impact, conditioned on observed evidence within the BN graph (§~\ref{subsec:method_posteriorprob})

    \For{$\text{num\_runs} = 1$ to $\text{max\_num\_runs}$}
        \For{$\text{num\_trials} = 1$ to $\text{max\_num\_trials}$}
            \State \textbf{Run} multi-objective optimisation to generate candidate countermeasure portfolios (§~\ref{subsec:method_multiobjoptimisation})
            
            \State \textbf{Update} BN graph mitigation attributes using top-performing candidate portfolio
        \EndFor

        \State \textbf{Identify} high-impact mitigation nodes for prioritisation under operational constraints, by analysing frequency-based heuristics of top-performing portfolios (§~\ref{subsec:method_freq_based_analysis})
    \EndFor
    \State
    \textbf{Recommend} final countermeasure portfolio 
\end{algorithmic}
\label{alg:framework_overview}
\end{algorithm}

\subsection{Bayesian Network Model Construction}
\label{subsec:method_bn_construction}

Extending the work of {\.Z}ebrowski et al. \cite{Zebrowski2022} and Bhosale et al. \cite{Bhosale2023,Bhosale2024}, we construct an abstract BN graph that captures the relevant cyber-physical structure of the target CPS. The construction process follows a systematic procedure:

\begin{itemize}
    \item \textbf{Node identification:} We define three classes of nodes: (i) \emph{asset nodes} representing physical or cyber components (e.g., PLCs, sensors, communication links), (ii) \emph{vulnerability nodes} representing exploitable weaknesses associated with assets, and (iii) \emph{hazard nodes} representing unsafe system states or adverse outcomes (e.g., loss of control, physical damage).
    \item \textbf{Dependency identification:} Directed edges are identified based on causal and conditional dependencies derived from system architecture, data or control flow diagrams, and expert knowledge. Specifically, edges are added from assets to their associated vulnerabilities, between vulnerabilities to represent multi-step attacks, and from compromised assets or vulnerabilities to hazards to capture failure propagation pathways.
    \item \textbf{Parametrisation:} Conditional probability tables are assigned to each node using a combination of sources, including historical incident data, vulnerability metrics (e.g., CVSS, EPSS, KEV scores), and expert judgement. When precise data is unavailable, qualitative likelihoods (i.e., "low", "medium", "high") are mapped to quantitative probabilities using predefined scales.
    \item \textbf{Model instantiation:} A generic BN graph structure is instantiated by integrating relevant assets, vulnerabilities, attack paths, and their associated parameters derived in the earlier steps. This step encodes attack entry points, attack paths, and attacker goals.
\end{itemize}

This BN graph serves as a shared structural backbone for both the \emph{Bayesian Attack Graph (BAG)} and the \emph{Bayesian Impact Graph (BIG)} used in subsequent stages of the framework. While the BAG models the likelihood of attack success based on the attack path connecting the attacker to vulnerabilities, hazards, and asset failures, the BIG models the likelihood of severe impact to the CPS using the same attributes within an identical BN graph. Given the fixed BN graph structure common to both the BAG and BIG, we incorporate scenario-specific assumptions and hypothesised attack paths to populate asset, vulnerability, and hazard nodes, and encode the resulting model in AutomationML to support systematic analysis and downstream optimisation.

\paragraph{\textbf{Running Example}}

\begin{figure}[ht]
    \centering
    \includegraphics[width=0.8\linewidth]{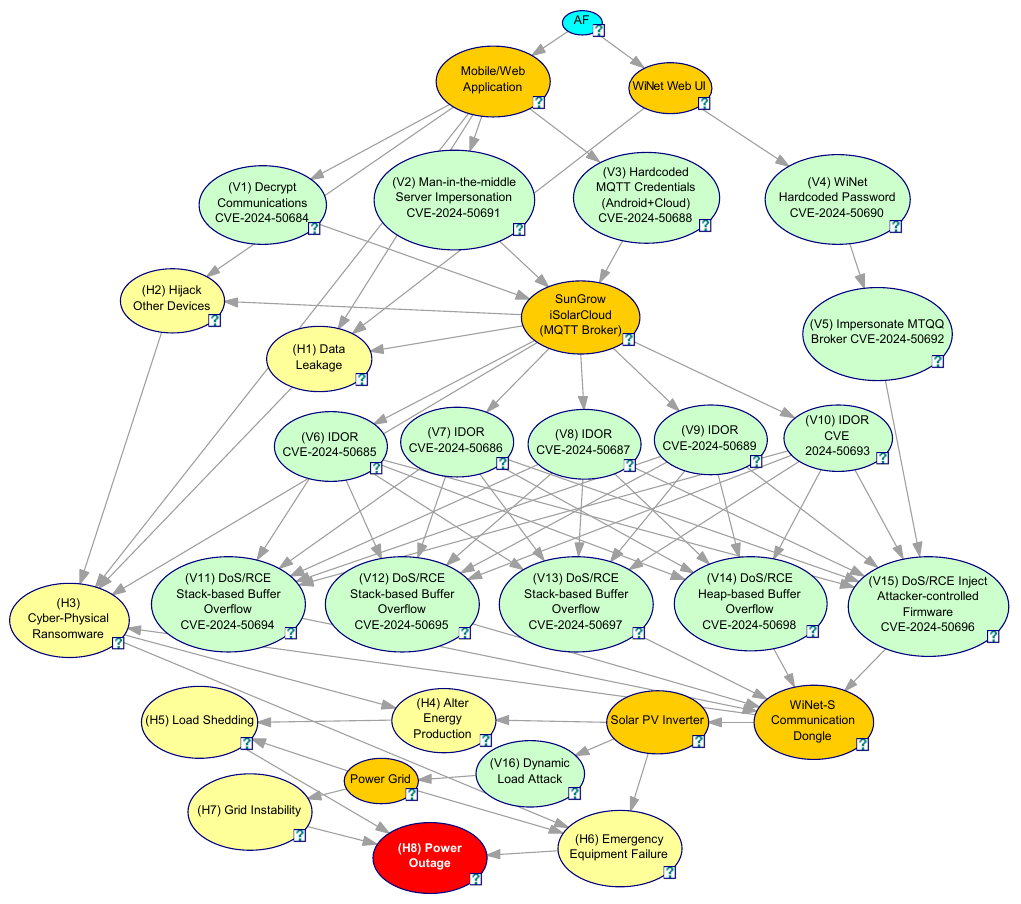}
    \caption{Bayesian Network graph of a hypothetical cyber-physical attack on solar PV inverters connected to a power grid. A high-resolution version of the graph is available in the project’s GitHub repository~\cite{Huang_GitHub}.}
    \label{fig:case_SolarPV_BN}
\end{figure}

Figure~\ref{fig:case_SolarPV_BN} shows a BN graph of a solar PV inverter attack scenario. The graph was constructed by mapping system topology and functional interdependencies referenced from the work by Dashevskyi et al.~\cite{Dashevskyi2025}. CPS asset, vulnerability, and hazard nodes are instantiated using information from CVE repositories alongside the attack scenario~\cite{Dabrowski2017}.

\subsection{Exposure Calculations}
\label{subsec:method_exposure_calculation}

This subsection outlines our methodology used to compute probabilities of attack success, which we term as \textit{exposure}, for vulnerability, asset, and hazard nodes within the BN graph.

\paragraph{Vulnerability Nodes}

The approach for vulnerability nodes needs to account for both publicly disclosed vulnerabilities with CVE identifiers, and proprietary or undocumented vulnerabilities lacking formal CVE enumeration. To calculate the exposures of CVE-linked vulnerability nodes, we use EPSS, LEV, and KEV metrics in conjunction with the \textit{maximal-envelope approach} introduced by Mell and Spring~\cite{Mell2025} to select the highest exposure probability among these complementary vulnerability scoring frameworks.

The Likely Exploited Vulnerability (LEV) metric is defined by Mell and Spring as:

\begin{equation}
\footnotesize
\text{LEV}(v, d_0, d_n) \geq 1 - \prod_{\forall d_i \in \text{dates}(d_0, d_n, 30)}
\bigl(1 - \text{epss}(v, d_i)
\times \text{weight}(d_i, d_n, 30)\bigr)
\label{eq:lev}
\end{equation}

\noindent where $v$ is the vulnerability; $d_0$ is the first date on which an EPSS score is available; $d_n$ is the date on which the calculation is performed (usually the present day); $\text{epss}(v,d)$ is the EPSS score for vulnerability $v$ on date $d$; and $\text{weight}(d_i,d_n,w)$ is the number of days between $d_i$ and $d_n$, limited by the window size $w$, divided by $w$.

The probability of exposure for a CVE-linked vulnerability, $P(E)_{\text{vuln, CVE}}$ is defined as:

\begin{equation}
P(E)_{\text{vuln, CVE}} = \max\left( \text{EPSS, LEV, KEV}\right)
\label{eq:p_exposure_vuln_cve}
\end{equation}
\noindent where $\text{KEV} \in \{0, 1\}$ acts as a logical ceiling, ensuring that any vulnerability confirmed to have been exploited in real-world scenarios is assigned a maximal exposure probability of 1; otherwise, it is assigned 0. If $KEV=1$, $P(E)_{\text{vuln,CVE}}$ becomes 1 regardless of EPSS or LEV, as this is a critical safety-conservative assumption for CPSs.

For proprietary or undocumented vulnerabilities lacking CVE identifiers, EPSS, LEV, and KEV metrics are unavailable for exposure probability calculation. Proxy CVSS scores are therefore derived from the CVSS base metrics, computed as follows:

\begin{equation}
    P(E)_{\text{vuln, CVSS}} = AV \times AC \times PR \times UI
\label{eq:p_exposure_vuln_noncve}
\end{equation}

\noindent where $AV$, $AC$, $PR$, and $UI$ represent the Attack Vector, Attack Complexity, Privileges Required, and User Interaction metrics, respectively. Each metric is assigned the corresponding numerical weight per the CVSS v3.1 specification~\cite{CVSS31}. Note that the CVSS v3.1 specification scales the product of these four metrics by a constant factor (8.22) when computing the exploitability sub-score. We omit this scaling constant deliberately as our purpose here is to treat the product as a pseudo-probability of exploitation, bounded naturally in [0, 1] since each metric weight is itself a probability-like value in that range.

From a CPS practitioner perspective, relying on uncalibrated CVSS scores introduces dangerous metric volatility and downstream risk miscalculations. In particular, CVSS scoring for zero-day or proprietary vulnerabilities is notoriously subjective and prone to initial assessor bias~\cite{Wunder2024}. To address this, we employ a \textit{Bayesian confidence calibration technique} inspired by Küppers et al.~\cite{Kuppers2021}. We model the observed proxy CVSS score as a distribution linked to a shared latent variable $P(E)^*_{\text{vuln}}$ with a variance of $\sigma^2_{\text{CVSS}}$, representing the unobserved true probability of exposure:

\begin{align}
P(E)_{\text{vuln, CVSS}} &\sim \mathcal{N}\left(P(E)^*_{\text{vuln}}, \sigma^2_{\text{CVSS}}\right) \label{eq:latent_distribution} 
\end{align}

Assuming a Gaussian prior as proposed by Knapik et al.\cite{Knapik2011}, where \( P(E)^*_{\text{vuln}} \sim \mathcal{N}(\mu_0, \sigma_0^2) \), the Bayesian latent variable formulation for the posterior parameters is thus represented as:

\begin{align}
    \mu_{\text{post}} &= \frac{
        \frac{\mu_0}{\sigma_0^2}
        + \frac{P(E)_{\text{vuln, CVSS}}}{\sigma^2_{\text{CVSS}}}
    }{
        \frac{1}{\sigma_0^2}
        + \frac{1}{\sigma^2_{\text{CVSS}}}
    }
    \label{eq:posterior_mean_nocve} \\
    \sigma^2_{\text{post}} &= \left(
        \frac{1}{\sigma_0^2}
        + \frac{1}{\sigma^2_{\text{CVSS}}}
    \right)^{-1}
    \label{eq:posterior_variance_nocve}
\end{align}

The resulting calibrated exposure probability estimate \( P(E)^*_{\text{vuln}} \sim \mathcal{N}(\mu_{\text{post}}, \sigma^2_{\text{post}}) \), serves as a unified, uncertainty-aware metric for supporting downstream risk assessment and mitigation optimisation within the BN graphs. Whilst \( \mu_{\text{post}} \) provides a calibrated point estimate for supporting downstream risk assessment, \( \sigma^2_{\text{post}} \)  serves as a supporting component available to practitioners for assessing the reliability of individual exposure estimates. In the current implementation, \( \sigma^2_{\text{post}} \) is surfaced as an uncertainty metric rather than formally propagated into subsequent stages of the pipeline, an extension identified for future work.

In addition, we incorporate an \textit{Attack Feasibility} (AF) modifier, inspired by the work of Xie et al.~\cite{Xie2010}, to ensure the formulation reflects real-world conditions more accurately. This modifier is applied to the probability calculations for all vulnerability nodes. Specifically, the AF modifier adjusts probability estimations by accounting for the system’s security posture, adversary capabilities, and environmental constraints. For instance, systems featuring strong defences and restricted access protocols yield lower AF values, reflecting a reduced likelihood of exploitation. Conversely, an unpatched CPS targeted by a sophisticated adversary results in a higher probability of a successful attack. The final probability of a successful attack for a given vulnerability node, irrespective of whether it has a CVE identifier, is defined as:

\begin{equation}
    P(A)_{\text{vuln}} = P(E)_{\text{vuln}} \times \phi
\label{eq:AF_modifier}
\end{equation}

\noindent where:
\begin{itemize}
    \small
    \item \( P(A)_{\text{vuln}} \) represents the probability that an attack of the vulnerability node is successful.
    \item \( P(E)_{\text{vuln}} \) denotes the probability of exposure for the vulnerability.
    \item \( \phi \) is the attack feasibility (AF) modifier, representing context-specific variables such as system defence posture, adversary capability, and environmental constraints.
\end{itemize}

As precise empirical data is generally unavailable in CPS contexts, a qualitative-to-quantitative mapping (Table~\ref{tab:af_mapping}) is applied to assign $\phi$ from a four-band scale corresponding to overall attack feasibility.

\begin{table}[ht]
\centering
\scriptsize
\caption{Qualitative-to-quantitative mapping for the AF modifier ($\phi$)}
\label{tab:af_mapping}
\begin{tabular}{@{}llp{5.5cm}@{}}
\toprule
\textbf{Band} & $\phi$ & \textbf{Criteria} \\
\midrule
Low & 0.25 & High exploitation difficulty and/or no demonstrated adversary capability against the target; exploitation requires significant resources, specialised access, or conditions unlikely to be met. \\
\addlinespace
Medium & 0.50 & Moderate exploitation difficulty; some enabling conditions present (e.g., partial access, known but unpatched weakness) without corroborating evidence of adversary intent or capability. \\
\addlinespace
Moderately high & 0.75 & Low-to-moderate exploitation difficulty, or credible but unconfirmed indicators of adversary capability/interest in the target. \\
\addlinespace
High & 1.00 & Low exploitation difficulty \emph{combined with} demonstrated real-world adversary capability against the target (e.g., confirmed active exploitation). \\
\bottomrule
\end{tabular}
\end{table}

\paragraph{Asset Nodes}

For an \textit{asset node}, the probability of exposure corresponds to the probability of asset failure because the node’s only failure event is its own failure, and is calculated using an exponential decay function, which models failure likelihood over time:

\begin{equation}
    P(E)_{\text{asset}} = 1 - e^{-\lambda t}
\label{eq:p_exposure_asset}
\end{equation}

\noindent where 
\( \lambda \) represents the failure rate per unit time, determining how frequently the asset is expected to degrade, and 
\( t \) denotes the operational duration, reflecting the asset’s time in service.

This formulation is grounded in Markov-based reliability analysis~\cite{Kumar2021}, in which asset failures are modelled as a stochastic process, capturing the progressive deterioration of system components over time. Through time-dependent degradation patterns, this approach provides a probabilistic estimation of failure likelihood, supporting predictive maintenance and risk-informed decision-making in CPS environments.

\paragraph{Hazard Nodes}

Finally, \textit{hazard nodes} represent hazardous events, failures, or disruptions influenced by attack propagation. As such, the occurrence of a hazard is influenced by the cyber-physical dependencies within the system. Specifically, downstream hazard exposure is certain ($P(E)=1$) once its parent hazard occurs or parent asset fails, as the failure propagates through the attack tree. The probability of exposure for a hazard node is thus defined as:

\begin{equation}
    P(E)_{\text{haz}} =
    \begin{cases} 
    1, & \text{if parent asset fails or parent hazard occurs} \\
    0, & \text{otherwise}
    \end{cases}
\label{eq:p_exposure_haz}
\end{equation}

We adopt deterministic hazard propagation as a safety-conservative choice, consistent with the worst-case reasoning underlying the KEV ceiling in Equation~\ref{eq:p_exposure_vuln_cve}. In practice, CPS environments often incorporate redundancy, protective relays, automated safety interlocks, and operator intervention, any of which may prevent propagation to a downstream node. However, reliably quantifying the failure probability of these protective mechanisms requires system-specific reliability data that is frequently unavailable or commercially sensitive, particularly for legacy CPS assets. In the absence of such data, assuming certain propagation ($P(E) = 1$) avoids the risk of understating hazard likelihood, which we consider the more operationally cautious failure mode for incident-response decision support. We acknowledge that this assumption will overstate risk in systems with effective protective layers, and we identify that probabilistic hazard propagation, conditioned on protection-system reliability where such data is available, is a natural extension once that data exists.

\paragraph{\textbf{Running Example}}

Continuing our running example of the solar PV inverter (Figure~\ref{fig:case_SolarPV_BN}), EPSS scores and KEV catalogue entries for CVE-identified vulnerabilities were retrieved from the FIRST EPSS and CISA databases respectively. Correspondingly, their LEV scores were computed per Equation~\ref{eq:lev} using EPSS scores over a 30-day window ($w=30$). For example, in the case of \texttt{V8} (CVE-2024-50687: iSolarCloud API allows unauthorised access to Sungrow device data and parameters), the corresponding vulnerability metrics are listed in Table~\ref{tab:vuln_exposure_attributes_PV}, yielding an exposure probability of $P(E)_{V8,CVE} = 0.0476$ (Equation~\ref{eq:p_exposure_vuln_cve}).

\begin{table}[ht]
\centering
\scriptsize
\caption{Vulnerability metrics for CVE-2024-50687 (retrieved 2026-06-18). Despite a high CVSS score (9.1), the LEV-derived exposure probability remains low (0.0476), indicating that although the vulnerability has a high theoretical severity score, it is unlikely to be exploited in real-world scenarios.}
\begin{tabular}{lp{5cm}}
\toprule
\textbf{Scoring System} & \textbf{Score} \\
\midrule
CVSS v3.1 & 9.1 \\
EPSS (Latest Score) & 0.0041 \\
LEV & 0.0476 (4.76\%) \\
KEV & 0 (Not listed in CISA catalogue) \\
\midrule
\(P(E)_{\text{V8,\,CVE}}\) & \textbf{0.0476}\\
\bottomrule
\end{tabular}
\label{tab:vuln_exposure_attributes_PV}
\end{table}

We also applied the AF modifier (Equation~\ref{eq:AF_modifier}) with a selected value of $\phi = 0.75$ to reflect the contextual conditions documented by Dashevskyi et al. in their report~\cite{Dashevskyi2025}. Specifically, the system's security posture was assessed as weak. Notably, the WiNet-S communication dongle exposed vulnerabilities that allowed device serial numbers to be harvested, contained hard-coded credentials that could be used to decrypt firmware updates, and accepted unauthenticated MQTT messages. These factors collectively enable remote code execution and persistent device takeover. Whilst the underlying vulnerabilities indicate a low barrier to exploitation, no real-world adversary has been observed exploiting these flaws to destabilise solar power grids, and all affected vendors have since released patches. Consequently, a moderately high AF modifier was adopted, reflecting low exploitation difficulty offset by the absence of demonstrated adversary capability and real-world attack validation.

The resulting probability of a successful attack on the vulnerability node is therefore:

\begin{equation}
P(A)_{V8, CVE} = P(E)_{V8, CVE} \times \phi = 0.0357
\end{equation}

In this example, \texttt{V16} is the only vulnerability node that does not have a CVE identifier, and represents a dynamic load-changing attack on the power grid. For this node, we approximated the CVSS-proxy vector based on information from the literature~\cite{Dabrowski2017}. The CVSS vector (\emph{CVSS:3.1/AV:N/AC:L/PR:N/UI:N/S:C/C:N/I:L/A:H}
) characterises a network-accessible attack within the CPS. AV:N indicates remote exploitation over a network; AC:L denotes low complexity requiring no specialised preconditions; PR:N and UI:N show that no privileges or user interaction are needed; S:C reflects changed scope impacting beyond the target system; C:N indicates no confidentiality loss; I:L shows low integrity compromise; while A:H signifies high availability impact. Using Equation~\ref{eq:p_exposure_vuln_noncve} with the corresponding CVSS metric values~\cite{CVSS31}, the fallback exposure probability estimate for \texttt{V16} is calculated as:
\begin{equation}
    P(E)_{V16, CVSS} = 0.85 \times 0.77 \times 0.85 \times 0.85 = 0.4729
\label{eq:p_exposure_v16}
\end{equation}

Next, we applied our Bayesian confidence calibration technique to evaluate the reliability of the fallback exposure probability estimate \( P(E)_{V16, CVSS} \), which is assumed to be low-confidence for illustrative purposes. For the prior distribution, we adopted a toy mean of $\mu_0 = 0.5$, reflecting the assumption that empirically similar vulnerabilities have a mean CVSS score of 0.5, together with a conservative standard deviation of $0.05$ (variance $\sigma_0^2 = 0.0025$). Finally, we selected a standard deviation of $0.20$ (variance $\sigma_{\text{CVSS}}^2 = 0.04$) to reflect the low confidence associated with the CVSS-proxy estimate. Substituting these values into Equations~\ref{eq:posterior_mean_nocve} and \ref{eq:posterior_variance_nocve} yields the following:

\begin{align}
\mu_{\text{post}} &= \frac{(0.5 / 0.0025) + (0.4729 / 0.04)}{(1 / 0.0025) + (1 / 0.04)} 
\approx \mathbf{0.4984} \\
\sigma^2_{\text{post}} &= \left( \frac{1}{0.0025} + \frac{1}{0.04} \right)^{-1}
= \mathbf{0.00235}
\end{align}

The posterior distribution of the shared latent vulnerability exposure probability \( P(E)^*_{V16, CVSS} \) is therefore given by:

\begin{equation}
P(E)^*_{V16, CVSS} \sim \mathcal{N}(0.4984, 0.00235)
\end{equation}

We performed a sensitivity analysis to evaluate this calibration, varying $\sigma_\text{CVSS}$ and observing the resulting shift in the posterior mean $\mu_\text{post}$ (Table~\ref{tab:sensitivity_analysis_solar}). Under low-confidence assumptions ($\sigma_\text{CVSS} \geq 0.20$), $\mu_\text{post}$ shifts by at most 0.30\% even as $\sigma_\text{CVSS}$ quadruples, confirming that the calibration is insensitive to imprecise uncertainty estimates when the prior is trusted. Under high-confidence assumptions ($\sigma_\text{CVSS} < 0.20$), the corresponding shift is larger (2.46\%), showing that assuming high confidence prematurely risks overstating the reliability of the proxy estimate.

\begin{table}[ht]
\centering
\scriptsize
\caption{Sensitivity analysis results (Solar PV inverter), showing the
effect of varying the CVSS-proxy standard deviation $\sigma_\text{CVSS}$ on the calibrated posterior estimate.}
\label{tab:sensitivity_analysis_solar}
\setlength{\tabcolsep}{3.5pt}
\begin{tabular}{llccccc}
\toprule
Scenario & $\mu_0$ & $\sigma_\text{CVSS}$ & $\sigma_\text{CVSS}^2$ & $\mu_{\text{post}}$ & $\sigma^2_{\text{post}}$ & $\Delta$ (\%) \\
\midrule
\multirow{4}{*}{\shortstack[l]{Solar PV\\(low conf.)}}
 & 0.5 & 0.20 & 0.04   & 0.4984 & 0.00235 & Baseline \\
 & 0.5 & 0.40 & 0.16   & 0.4996 & 0.00246 & 0.24\%  \\
 & 0.5 & 0.60 & 0.36   & 0.4998 & 0.00248 & 0.28\%  \\
 & 0.5 & 0.80 & 0.64   & 0.4999 & 0.00249 & 0.30\%  \\
\midrule
\multirow{4}{*}{\shortstack[l]{Solar PV\\(high conf.)}}
 & 0.5 & 0.05 & 0.0025 & 0.4865 & 0.00125 & Baseline \\
 & 0.5 & 0.10 & 0.0100 & 0.4946 & 0.00200 & 1.67\%   \\
 & 0.5 & 0.15 & 0.0225 & 0.4973 & 0.00225 & 2.23\%   \\
 & 0.5 & 0.20 & 0.0400 & 0.4984 & 0.00235 & 2.46\%   \\
\bottomrule
\end{tabular}
\end{table}

Finally, the AF modifier of $\phi = 0.75$ was applied. The resulting probability of a successful attack on the vulnerability node \texttt{V16} is therefore:

\begin{equation}
P(A)_{V16, CVSS} = P(E)^*_{V16, CVSS} \times \phi = 0.3738
\end{equation}

Other than vulnerability nodes, exposure probabilities were also assigned to asset and hazard nodes. For asset nodes, Equation~\ref{eq:p_exposure_asset} was applied using an operational duration defined from 1 January 2024 to the current date. Failure rates for individual assets, such as the WiNet-S communication dongle connected to the inverter, were estimated from published literature~\cite{Bhosale2023} and the technical specifications of comparable components sourced from publicly available documentation. Exposure probabilities for hazard nodes were subsequently assigned using Equation~\ref{eq:p_exposure_haz}, completing the probabilistic exposure attributes of all node types within the BN graph.

\subsection{Impact Calculations}
\label{subsec:method_impact_calculation}

This subsection outlines our methodology used to compute impact-related attributes within the BN graph: the probability of severe impact for vulnerability nodes, and the structural impact score for asset and hazard nodes.

\paragraph{Vulnerability Nodes}

In the case of vulnerability nodes (both CVE- and non-CVE-linked), we use the following formulation, drawing on the corresponding CVSS metric values~\cite{CVSS31}, to compute the probability of severe impact to the CPS:

\begin{equation}
    P(\text{Impact})_{\text{vuln}} = 1 - \left[(1 - C) \cdot (1 - I) \cdot (1 - A)\right]
\label{eq:p_impact_vuln}
\end{equation}

\noindent where:
\begin{itemize}
    \small
    \item \( P(\text{Impact})_{\text{vuln}} \) denotes the probability of severe impact associated with the vulnerability node.
    \item \( C, I, A \) represent the confidentiality, integrity, and availability impact metrics, respectively. For vulnerabilities that do not have CVE identifiers, the impact metrics are estimated based on available system context, attack surface semantics, or analogous vulnerabilities with similar functionality.
\end{itemize}

\paragraph{Asset and Hazard Nodes}

In the case of asset and hazard nodes, we compute a structural impact score that reflects the number of child nodes each individual node directly influences, capturing its potential to propagate disruption or failure through the system \cite{Akoglu2015,Bhosale2023}. This is formulated as:

\begin{equation}
    S(a) = \frac{\text{No. of connected child nodes}}{\text{Total no. of nodes in the BN graph}}
\label{eq:structural_impact_score}
\end{equation}

In this formulation, the structural impact score lies within the interval [0, 1], and asset or hazard nodes with greater downstream reach are assigned proportionally higher scores. We do not, however, interpret $ S(a) $ as a validated severity measure: a relay node with a single downstream hazard link but severe consequence could score lower than a node connected to several low-consequence assets, even though the former carries the greater actual risk. We leave severity-weighted or path-based propagation scoring as a refinement for subsequent work.

\paragraph{\textbf{Running Example}}

In our running example, the impact probabilities of vulnerability nodes were calculated using Equation~\ref{eq:p_impact_vuln}, drawing on the confidentiality, integrity, and availability metric values derived from their respective CVSS vectors. For instance, the vulnerability node \texttt{V8} (CVE-2024-50687) has a CVSS vector: \emph{CVSS:3.1/AV:N/AC:L/PR:N/UI:N/S:U/C:H/I:H/A:N}. Here, C:H indicates high confidentiality loss (metric value: 0.56); I:H shows high integrity compromise (metric value: 0.56); and A:N signifies no availability impact (metric value: 0). Accordingly, the impact probability for \texttt{V8} was calculated as follows:

\begin{equation}
\footnotesize
P(Impact)_{V8} = 1 - [(1-0.56)\times(1-0.56)\times(1-0)] = 0.8064
\end{equation}

Within the solar PV inverter BN graph presented in Figure~\ref{fig:case_SolarPV_BN}, which comprises 30 nodes in total (i.e., excluding the Attacker (AF) node), the impact probabilities of hazard and asset nodes were derived from the underlying interdependencies among them. For illustration, successful exploitation of an asset node \textit{app}, specifically the Sungrow iSolarCloud mobile/web application, directly influences three vulnerability nodes: \texttt{V1}, \texttt{V2}, and \texttt{V3}. Using Equation~\ref{eq:structural_impact_score}, this asset node's structural impact score was calculated as follows:

\begin{equation}
S(app) = \frac{3}{30} = 0.1
\end{equation}

\subsection{Posterior Probability Calculations}
\label{subsec:method_posteriorprob}

Once the exposure probabilities and impact-related attributes for vulnerability, asset and hazard nodes are calculated, the posterior probabilities of successful attack and of severe impact to the CPS, given the evidence of a vulnerability node being exploited, for example, can be derived using conditional probability tables and variable elimination~\cite{Zhang1996}. We note that the posterior probability of severe impact, denoted $ P(I) $ below, incorporates a combination of severity-grounded (Equation~\ref{eq:p_impact_vuln}) and structural (Equation~\ref{eq:structural_impact_score}) inputs. Where the attack path traverses an asset or hazard node, $ P(I) $ incorporates the structural impact score discussed in Section~\ref{subsec:method_impact_calculation}.

\paragraph{\textbf{Running Example}}

For the running example, the joint conditional probabilities for exposure and impact associated with the attacker's goal are shown in Table~\ref{tab:cpt_joint_solarpv}. These values were derived from Bayesian inference computations based on estimates, posterior parameters and updated evidence from the risk model.

\begin{table}[ht]
\centering
\caption{Joint Conditional Probability Table for Exposure and Impact of H8\_Power\_Outage (State (0): Successful attack)}
{
\begin{tabular}{lcc}
\toprule
State & $P(E)$ & $P(I)$ \\
\midrule
\texttt{(0)} & 0.4706 & 0.9153 \\
\texttt{(1)} & 0.5294 & 0.0847 \\
\bottomrule
\end{tabular}
}
\label{tab:cpt_joint_solarpv}
\end{table}

The results show a posterior likelihood of successful attack (exposure) of $P(E) = 0.4706$, and a posterior probability of severe CPS impact given that the attack succeeds of $P(I|E) = 0.9153$ (denoted $P(I)$ for brevity throughout). The product of these two probabilities yields a composite risk score of $R = P(E) \times P(I) = 43.07\%$, indicating a moderate risk of CPS disruption in this running example.

\subsection{Multi-Objective Optimisation}
\label{subsec:method_multiobjoptimisation}

Our framework employs multi-objective optimisation to assess and weigh trade-offs between \textit{countermeasure portfolios}, \textit{failure probabilities} and \textit{impact ratings} to inform consequence-driven and risk-informed decision support in CPSs.

\paragraph{Countermeasure Portfolios}

Countermeasure portfolios are instantiated as tuples of \textit{Probability of Mitigation} attributes, embedded within each vulnerability node in the BAG and BIG models. Each portfolio is defined as an ordered tuple $$\mathbf{M} = \left(P(M_1), P(M_2), \dots, P(M_n)\right)$$ comprising the mitigation probabilities $P(M_i)$ associated with each vulnerability $V_i\ (i = 1, 2, \dots, n)$ in the BAG model. A higher mitigation probability implies a lower likelihood of successful exploitation at the corresponding node. The ordering of nodes within these tuples reflects the relative effectiveness of mitigation strategies, thereby guiding prioritisation under operational constraints.

\paragraph{Failure Probabilities}

The first author's professional experience suggests that asset failure probabilities can be influenced by mitigation-induced risks. For instance, the application of faulty or unstable security patches to a CPS asset may introduce unintended consequences, including asset failure. This complexity is exacerbated when an asset is linked to multiple vulnerabilities, each associated with distinct mitigation-related risks. In our framework, each countermeasure is associated with a risk adjustment function that governs its net contribution to asset-level failure. Accordingly, the failure probability of a CPS asset is defined as a bounded, risk-adjusted function of the mitigations applied across its associated vulnerabilities:

\begin{equation}
    P(F_a \mid \mathbf{M}) = \min\left(1.0,\ \frac{\sum_{i \in \mathcal{V}_a} P(\text{Fail}_{\text{mit}}^i)}{|\mathcal{V}_a|}\right)
    \label{eq:risk_adjustment}
\end{equation}

\noindent
where:
\begin{itemize}
    \small
    \item $P(F_a \mid \mathbf{M})$ is the probability of failure for CPS asset $a$ given the countermeasure portfolio $\mathbf{M}$;
    \item $\mathcal{V}_a$ is the set of vulnerabilities associated with asset $a$;
    \item $P(\text{Fail}_{\text{mit}}^i)$ is the probability that mitigation applied to vulnerability $i$ causes unintended adverse consequences that contribute to asset failure;
    \item The $\min$ function ensures the total failure probability remains within the interval $[0, 1]$.
\end{itemize}

In practice, failure probabilities specific to each vulnerability mitigation should be corroborated with field data, such as user- or vendor-reported issues with deployed mitigations (e.g., software patches for the corresponding vulnerabilities). In the absence of such data for this study, a common, neutral value of $P(\text{Fail}_{\text{mit}}) = 0.5$ was adopted for illustrative purposes.

\paragraph{Impact Ratings}

Another important element in the framework is the \textit{Impact Rating} attribute, which is embedded in the derivation of countermeasure portfolios. By acting as a modifier for mitigation probabilities, the impact rating tailors mitigation strategies to address the multifaceted consequences of successful attacks across safety, operational, financial and informational domains. It is computed by adapting the formulation proposed by Amro et al.~\cite{Amro2023}:

\begin{equation}
    \text{Impact Rating} = \frac{\sum_{j \in \{S, F, I, O, St\}} \left( \text{Factor}_{j} \cdot \text{Criticality}_{j} \right)}{\sum_{j \in \{S, F, I, O, St\}} \text{Factor}_{j}}
    \label{eq:impact_rating}
\end{equation}

\noindent where:
\begin{itemize}
    \small
    \item \( \text{S} \): safety impact — refers to potential harm to human life or physical injury resulting from a CPS failure or security breach.
    \item \( \text{F} \): financial impact — quantifies monetary losses due to CPS equipment damage, service downtime or disrupted business operations.
    \item \( \text{I} \): informational impact — relates to the compromise, loss or exposure of sensitive data and system configurations.
    \item \( \text{O} \): operational impact — reflects the degree to which core CPS functions or processes are degraded or interrupted.
    \item \( \text{St} \): staging impact — indicates the ability to enable further attacks or facilitate persistence, privilege escalation or lateral movement.
    \item \( \text{Factor}_{j} \): the relative weight assigned to impact type \( j \).
    \item \( \text{Criticality}_{j} \): the criticality level associated with impact type \( j \).
\end{itemize}

The optimisation process begins by randomly generating countermeasure portfolios, denoted as $$\mathbf{M}^{(j)} \ (j = 1, 2, \dots, n)$$ Each portfolio’s mitigation values are encoded into the BN graph and used to infer the posterior probabilities of both attack success and severe impact on the CPS, via Bayesian inference algorithms. These probabilistic outcomes are treated as parameters within a multi-objective search space. Pareto-optimal portfolios are then identified using the Optuna library~\cite{Optuna2019}, balancing the competing objectives of minimising the likelihood of attack success, and at the same time, minimising the probability of severe impact and maximising system availability. The resulting Pareto fronts form the foundation of our proposed decision support mechanism.

\paragraph{System Availability}

System availability is the third optimisation objective alongside minimising the likelihoods of attack success and severe impact. We define availability under a candidate countermeasure portfolio $\mathbf{M}$ as the Impact-Rating-weighted complement of asset failure probability, using the per-component Impact Rating already computed via Equation~\ref{eq:impact_rating} as the criticality weight for each asset's mitigation-adjusted failure probability from Equation~\ref{eq:risk_adjustment}:

\begin{equation}
    \text{Availability} = 1 - \frac{\sum_{a \in Assets} ( \text{Impact Rating}_{a} \cdot P(F_a \mid \mathbf{M}))} {\sum_{a \in Assets} \text{Impact Rating}_{a}}
    \label{eq:availability}
\end{equation}

\noindent
where \( P(F_a \mid \mathbf{M}) \) is the failure probability of CPS asset $a$ under portfolio $\mathbf{M}$ and $\text{Impact Rating}_{a}$ is the corresponding component's impact rating. Weighting by Impact Rating ensures that the availability objective penalises the failure of operationally-critical assets more heavily than that of lower-criticality assets, consistent with the treatment of impact severity elsewhere in the framework. By construction, this formulation is bounded in $[0, 1]$, since \( P(F_a \mid \mathbf{M}) \) is itself bounded in $[0, 1]$ and the weights form a normalised average.

\paragraph{\textbf{Running Example}}

In our running example, Equation~\ref{eq:risk_adjustment} was used to calculate the risk-adjusted failure probabilities of CPS assets such as the WiNet-S communication dongle, which are connected to multiple vulnerability nodes, each associated with distinct mitigation-related risks.

\begin{table}[ht]
\centering
\tiny
\caption{Impact Ratings for Solar PV Inverter Components}
\label{tab:impact_attributes_solarpv}
\begin{tabular}{lp{0.1cm}p{0.1cm}p{0.1cm}p{0.1cm}p{0.1cm}p{0.1cm}p{0.1cm}p{0.1cm}p{0.1cm}p{0.1cm}c}
\toprule
Component & $F_S$ & $C_S$ & $F_F$ & $C_F$ & $F_I$ & $C_I$ & $F_O$ & $C_O$ & $F_{St}$ & $C_{St}$ & Impact Rating \\
\midrule
Mobile/Web App & 1 & 0 & 0.25 & 0 & 0.25 & 0.5 & 0.75 & 0 & 0.5 & 0.25 & 0.0909 \\
WiNet Web      & 1 & 0 & 0.25 & 0 & 0.25 & 0.5 & 0.75 & 0 & 0.5 & 0.25 & 0.0909 \\
MQTT Broker    & 1 & 0 & 0.25 & 0 & 0.25 & 0.5 & 0.75 & 0.5 & 0.5 & 0.5 & 0.2727 \\
WiNet-S Dongle & 1 & 0 & 0.25 & 0 & 0.25 & 0.5 & 0.75 & 0.5 & 0.5 & 0.5 & 0.2727 \\
PV Inverter    & 1 & 0 & 0.25 & 0.25 & 0.25 & 0.75 & 0.75 & 0.75 & 0.5 & 0.5 & 0.3864 \\
Power Grid     & 1 & 0.5 & 0.25 & 0.75 & 0.25 & 0 & 0.75 & 1 & 0.5 & 0.75 & 0.6591 \\
\bottomrule
\end{tabular}
\end{table}

In addition, Equation~\ref{eq:impact_rating} was used to calculate impact ratings with specific weights and criticality values corresponding to the CPS context, summarised in Table~\ref{tab:impact_attributes_solarpv}. The impact rating of the solar PV inverter was calculated to be 0.3864, reflecting a maximum safety factor of 1.0, based on the overarching principle that safety must never be compromised, alongside more moderate financial, informational, operational, and staging weightings that scale with each factor's practical consequence. These impact ratings were then multiplied by the corresponding mitigation probabilities to derive the countermeasure portfolio.

\begin{figure}[ht]
    \centering
    \includegraphics[width=0.75\linewidth]{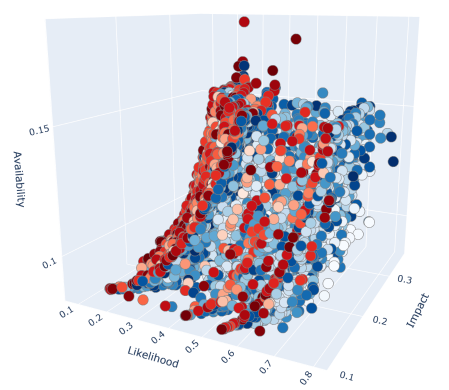}
    \caption{Pareto fronts derived over 10,000 multi-objective optimisation trials in the solar PV inverter case study}
    \label{fig:case_SolarPV_Plot_3D}
\end{figure}

Finally, a multi-optimisation study was performed to identify Pareto-optimal countermeasure portfolios. The scatter plot (Figure~\ref{fig:case_SolarPV_Plot_3D}) shows a curved, fairly smooth slope across the likelihood and impact axes, rising as impact increases. This points to a consistent relationship between the two dimensions, wherein higher-impact scenarios tend to coincide with higher availability. The smoothness of the slope suggests the CPS environment has underlying systemic regularities, likely driven by the causal structure captured in the BN graph. Between approximately 0.25 and 0.30 on the \textit{Impact} axis, the plot also flattens into a plateau, indicating that some countermeasure portfolios reduce the likelihood of severe impact without meaningfully improving system availability.

\subsection{Frequency-Based Heuristic Analysis}
\label{subsec:method_freq_based_analysis}

Lastly, a frequency-based heuristic analysis of Pareto-optimal portfolios is conducted across multiple optimisation runs to rank and identify the vulnerability nodes that most consistently and effectively contributed to high-performing countermeasure configurations. These nodes are strong candidates for prioritisation in mitigation strategies, particularly under operational constraints during incident response.

\paragraph{\textbf{Running Example}}

As shown in Figure~\ref{fig:case_SolarPV_freq_af}, vulnerabilities \texttt{V1}, \texttt{V2}, \texttt{V3}, and \texttt{V15} consistently demonstrate high mitigation effectiveness across portfolio instances. The underlying BN graph structure (Figure~\ref{fig:case_SolarPV_BN}) corroborates this finding, highlighting these nodes as critical entry points for downstream attack propagation in the CPS. Furthermore, sensitivity analysis across varying AF settings shows that while the Pareto-optimal configuration remains stable, several vulnerability nodes exhibit larger rank variance, suggesting that their relative mitigation effectiveness is dependent on how feasible the attack is.

\begin{figure}[ht]
\centering
\begin{tikzpicture}
\begin{axis}[
    width=\linewidth,
    height=4cm,
    ymin=2, ymax=13,
    ylabel={Average Rank Position},
    xlabel={Vulnerability Nodes},
    symbolic x coords={V1,V2,V3,V4,V5,V6,V7,V8,V9,V10,V11,V12,V13,V14,V15,V16},
    xtick=data,
    xticklabel style={rotate=60, anchor=east, font=\tiny},
    tick label style={font=\tiny},
    label style={font=\scriptsize},
    grid=both,
    grid style={line width=.1pt, draw=gray!20},
    major grid style={line width=.2pt, draw=gray!40},
    legend style={
        at={(0.5,1.05)},
        anchor=south,
        legend columns=4,
        font=\tiny,
        /tikz/every even column/.append style={column sep=0.2cm}
    },
]

\addplot[color=blue, mark=*, mark size=1.2pt, thick] coordinates {
    (V1,5.83) (V2,5.25) (V3,5.25) (V4,10.75) (V5,10.53) (V6,7.85) (V7,9.49) (V8,9.26) 
    (V9,8.15) (V10,8.73) (V11,10.79) (V12,10.85) (V13,10.61) (V14,10.42) (V15,4.05) (V16,8.19)
};

\addplot[color=red, mark=square*, mark size=1.2pt, thick] coordinates {
    (V1,6.34) (V2,5.81) (V3,5.37) (V4,10.95) (V5,10.74) (V6,9.10) (V7,9.16) (V8,8.69) 
    (V9,7.58) (V10,8.96) (V11,10.83) (V12,10.11) (V13,10.18) (V14,10.46) (V15,3.77) (V16,7.95)
};

\addplot[color=green!60!black, mark=triangle*, mark size=1.4pt, thick] coordinates {
    (V1,6.7) (V2,4.9) (V3,6.15) (V4,10.6) (V5,10.1) (V6,8.7) (V7,8.3) (V8,7.55) 
    (V9,10.1) (V10,9.7) (V11,10.7) (V12,10.1) (V13,10.3) (V14,11.7) (V15,3.35) (V16,7.05)
};

\addplot[color=orange!90!black, mark=diamond*, mark size=1.4pt, thick] coordinates {
    (V1,6.17) (V2,5.41) (V3,4.88) (V4,11.07) (V5,10.84) (V6,9.44) (V7,8.86) (V8,8.64) 
    (V9,8.31) (V10,9.34) (V11,10.69) (V12,10.02) (V13,10.61) (V14,10.13) (V15,3.99) (V16,7.6)
};

\legend{AF=0.25, AF=0.5, AF=0.75, AF=1.0}

\end{axis}
\end{tikzpicture}
\caption{Average rank positions of mitigation probabilities across 100 runs of 10,000 optimisation trials, repeated across varying AF settings.}
\label{fig:case_SolarPV_freq_af}
\end{figure}

In summary, the key findings of the running example of a solar PV inverter attack are:
\begin{itemize}
  \item Optimisation results show systemic regularity, suggesting underlying causal structures within the BN graph.
  \item A subset of countermeasure portfolios reduce the likelihood of severe impact, yet offer limited gains in system availability.
  \item Vulnerabilities that demonstrate consistently high mitigation effectiveness reflect their positions as critical entry points in the BN graph.
  \item While Pareto-optimal countermeasure portfolios remain stable across varying AF settings, individual node ranks display varying degrees of sensitivity to AF selection.
\end{itemize}

\section{Further Evaluation Across CPS Scenarios}
\label{sec:evaluation}

This section evaluates the practical utility and applicability of our proposed framework by applying the methodology to another two representative CPS scenarios beyond the initial solar PV inverter case study. The corresponding results are presented and analysed to highlight key insights, limitations, and implications for real-world deployment.

\subsection{Ukrainian Power Grid (BlackEnergy) Attack}
\label{subsec:cs_blackenergy}

\begin{figure}[t]
    \centering
    \includegraphics[width=0.8\linewidth]{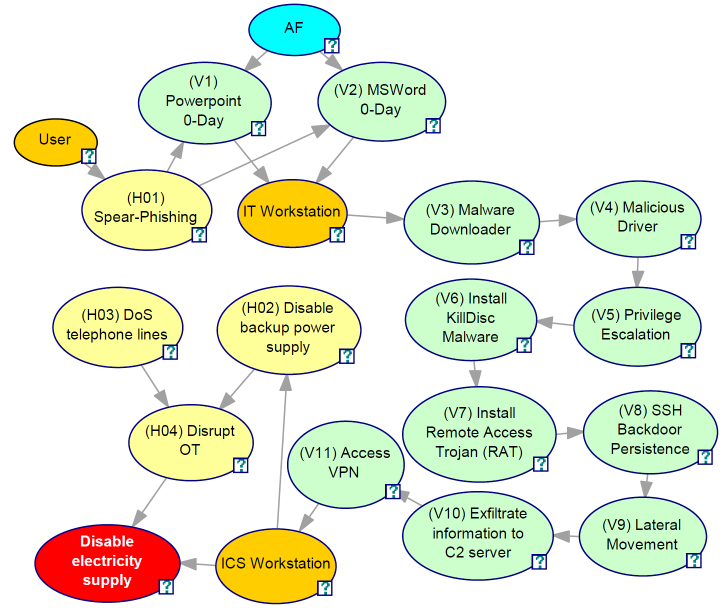}
    \caption{Bayesian Network of the Ukrainian power grid (BlackEnergy) attack. A high-resolution version is available in the project’s GitHub repository (\cite{Huang_GitHub}).}
    \label{fig:case_BlackEnergy_BN}
\end{figure}

Our proposed framework was applied to enable adaptive mitigation of the 2015 Ukrainian power grid attack, which notoriously leveraged the BlackEnergy malware to attack a CPS~\cite{BlackEnergy2016}. Figure~\ref{fig:case_BlackEnergy_BN} presents the corresponding Bayesian Network graph of the attack tree~\cite{Kumar2022} created with the GeNIe Modeler tool~\cite{Genie2025}. Based on this structure, the BN graph is constructed using available information to define the asset, vulnerability, and hazard nodes, and encoded in AutomationML.

Next, we calculated the exposure and impact probability values using the methodology outlined in Section~\ref{sec:methodology}. Two of the vulnerability nodes were listed in the National Vulnerability Database (NVD) \cite{NVD2026} with CVE identifiers, while the rest were not. Table~\ref{tab:vuln_exposure_attributes_BE} lists the vulnerability metrics for vulnerability node \texttt{V2} (CVE-2014-1761: Microsoft Word Remote Code Execution), yielding an exposure probability of $P(E)_{V2,CVE} = 1.0$ (Equation~\ref{eq:p_exposure_vuln_cve}) because it was listed in the CISA catalogue of known exploited vulnerabilities.

\begin{table}[t]
\centering
\scriptsize
\caption{Vulnerability metrics for CVE-2014-1761 (retrieved 2026-06-18). Despite age ($>$10 years), high CVSS (7.8), LEV (1.0), and KEV listing yield maximum exposure probability (1.0) -- indicating the vulnerability is almost certain to be exploited in real-world scenarios.}
\begin{tabular}{lp{5cm}}
\toprule
\textbf{Scoring System} & \textbf{Score} \\
\midrule
CVSS v3.1 & 7.8 \\
EPSS (Latest Score) & 0.7773 \\
LEV & 1.000 (100\%) \\
KEV & 1 (Listed in CISA catalogue) \\
\midrule
$P(E)_{\text{V2, CVE}}$ & \textbf{1.0}\\ 
\bottomrule
\end{tabular}
\label{tab:vuln_exposure_attributes_BE}
\end{table}

The AF modifier (Equation~\ref{eq:AF_modifier}) was then applied. A value of $\phi = 1.0$ was selected to reflect the combination of confirmed exploitation and the contextual conditions documented in the SANS/E-ISAC post-incident analysis \cite{BlackEnergy2016}. Specifically, the system's security posture was assessed as very weak. The initial attack vector exploited corporate workstations with macro-enabled Office documents, VPN access lacked two-factor authentication, and there was no network segmentation between the IT and OT environments, nor any resident capability to monitor the ICS network for anomalous activity. Compounding this, the adversary demonstrated exceptional capability, comprising a highly skilled, well-resourced team that conducted over six months of undetected reconnaissance before executing a highly synchronised, multistage attack. The resulting probability of a successful attack on the vulnerability node \texttt{V2} is therefore \( P(A)_{V2, CVE} = P(E)_{V2, CVE} \times \phi = 1.0 \).

Next, the exposure probabilities of non-CVE vulnerabilities were estimated and calibrated. Taking the example of one such vulnerability node \texttt{V3}, which represents a malware downloader, the CVSS-proxy vector was approximated as: \emph{CVSS:3.1/AV:L/AC:L/PR:N/UI:N/S:U/C:H/I:H/A:H}.

This CVSS vector characterises a local attack scenario, interpreted per the metric definitions in Section~\ref{subsec:method_exposure_calculation}, and reflects high confidentiality, integrity, and availability loss should the vulnerability be exploited. Based on the CVSS vector, the fallback exposure probability estimate was derived as \( P(E)_{V3,CVSS} = 0.306 \).

We calibrated this estimate using Equation~\ref{eq:posterior_mean_nocve}, arriving at a likelihood variance of $\sigma_{\text{CVSS}}^2 = 0.0025$ (standard deviation $0.05$), reflecting high confidence in the fallback estimate for illustrative purposes. For the prior distribution, we adopted a toy mean of $\mu_0 = 0.3$, assumed as an empirical mean for similar vulnerabilities, together with a variance of $\sigma_0^2 = 0.0025$ (standard deviation $0.05$). Using the Bayesian latent variable formulation (Equation~\ref{eq:posterior_mean_nocve}), the posterior distribution is thus:

\begin{align}
\mu_{\text{post}} &= \frac{(0.3 / 0.0025) + (0.306 / 0.0025)}{(1 / 0.0025) + (1 / 0.0025)} 
\approx \mathbf{0.303} \\
\sigma^2_{\text{post}} &= \left( \frac{1}{0.0025} + \frac{1}{0.0025} \right)^{-1}
= \mathbf{0.00125}
\end{align}

The posterior distribution of the shared latent exposure probability \( P(E)^*_{V3, CVSS} \) is therefore given by  \( P(E)^*_{V3, CVSS} \sim \mathcal{N}(0.303, 0.00125) \). The low variance indicates that the posterior calibrated estimate can be used with relatively high confidence within the Bayesian network. Next, we performed a sensitivity analysis over two prior distributions ($\mu_0 = 0.3$, $\mu_0 = 0.5$ -- reflecting a prior that the proxy estimate of $0.306$ disagrees with). Comparing the results in Table~\ref{tab:sensitivity_analysis_blackenergy}, the $\mu_0 = 0.5$ prior distribution shows a markedly large change from the baseline (21.24\%) when $\sigma_{\text{CVSS}}$ is increased to $0.20$. By contrast, the $\mu_0 = 0.3$ distribution is relatively stable, showing a maximum deviation of only 0.87\%. We discuss this further in Section~\ref{sec:discussion}.

\begin{table}[ht]
\centering
\scriptsize
\caption{Sensitivity analysis results (BlackEnergy), showing the
effect of varying the CVSS-proxy standard deviation $\sigma_\text{CVSS}$ on the calibrated posterior estimate across two priors.}
\label{tab:sensitivity_analysis_blackenergy}
\setlength{\tabcolsep}{3.5pt}
\begin{tabular}{llccccc}
\toprule
Scenario & $\mu_0$ & $\sigma_\text{CVSS}$ & $\sigma_\text{CVSS}^2$ & $\mu_{\text{post}}$ & $\sigma^2_{\text{post}}$ & $\Delta$ (\%) \\
\midrule
\multirow{4}{*}{\shortstack[l]{BlackEnergy\\($\mu_0=0.3$)}}
 & 0.3 & 0.05 & 0.0025 & 0.3030 & 0.00125 & Baseline \\
 & 0.3 & 0.10 & 0.0100 & 0.3012 & 0.00200 & -0.59\%  \\
 & 0.3 & 0.15 & 0.0225 & 0.3006 & 0.00225 & -0.79\%  \\
 & 0.3 & 0.20 & 0.0400 & 0.3004 & 0.00235 & -0.87\%  \\
\midrule
\multirow{4}{*}{\shortstack[l]{BlackEnergy\\($\mu_0=0.5$)}}
 & 0.5 & 0.05 & 0.0025 & 0.4030 & 0.00125 & Baseline \\
 & 0.5 & 0.10 & 0.0100 & 0.4612 & 0.00200 & 14.44\%   \\
 & 0.5 & 0.15 & 0.0225 & 0.4806 & 0.00225 & 19.26\%   \\
 & 0.5 & 0.20 & 0.0400 & 0.4886 & 0.00235 & 21.24\%   \\
\bottomrule
\end{tabular}
\end{table}

The AF modifier was similarly applied with $\phi = 1.0$. The resulting probability of a successful attack on the vulnerability node \texttt{V3} is therefore \(
P(A)_{V3, CVSS} = P(E)^*_{V3, CVSS} \times \phi = 0.3004 \).

\begin{table}[ht]
\centering
\caption{Joint Conditional Exposure and Impact Probabilities (BlackEnergy)}
{
\begin{tabular}{lcc}
\toprule
State & $P(E)$ & $P(I)$ \\
\midrule
\texttt{(0)} & 0.1640 & 0.2393 \\
\texttt{(1)} & 0.8360 & 0.7607 \\
\bottomrule
\end{tabular}
}
\label{tab:cpt_joint_blackenergy}
\end{table}

Table~\ref{tab:cpt_joint_blackenergy} lists the conditional probabilities for exposure and impact associated with the attacker’s goal, derived using Bayesian inference via variable elimination. The results show a posterior likelihood of successful attack of $P(E) = 0.1640$, and a corresponding posterior probability of severe impact on the CPS of $P(I) = 0.2393$. The composite risk score is computed as $R = P(E) \times P(I) = 3.92\%$ in the BlackEnergy attack. As with the running example, $P(I)$ here incorporates a combination of severity-grounded and structural inputs. Whilst this score may appear low given that this scenario is grounded in a confirmed real-world incident, it does not contradict the historical record. Rather, it reflects the framework's modelling of attack success likelihood under the specified security posture, wherein propagation of posterior probabilities across the multi-stage attack chain suppresses the marginal exposure probability at the attacker goal node.

\begin{figure}[ht]
    \centering
    \includegraphics[width=0.75\linewidth]{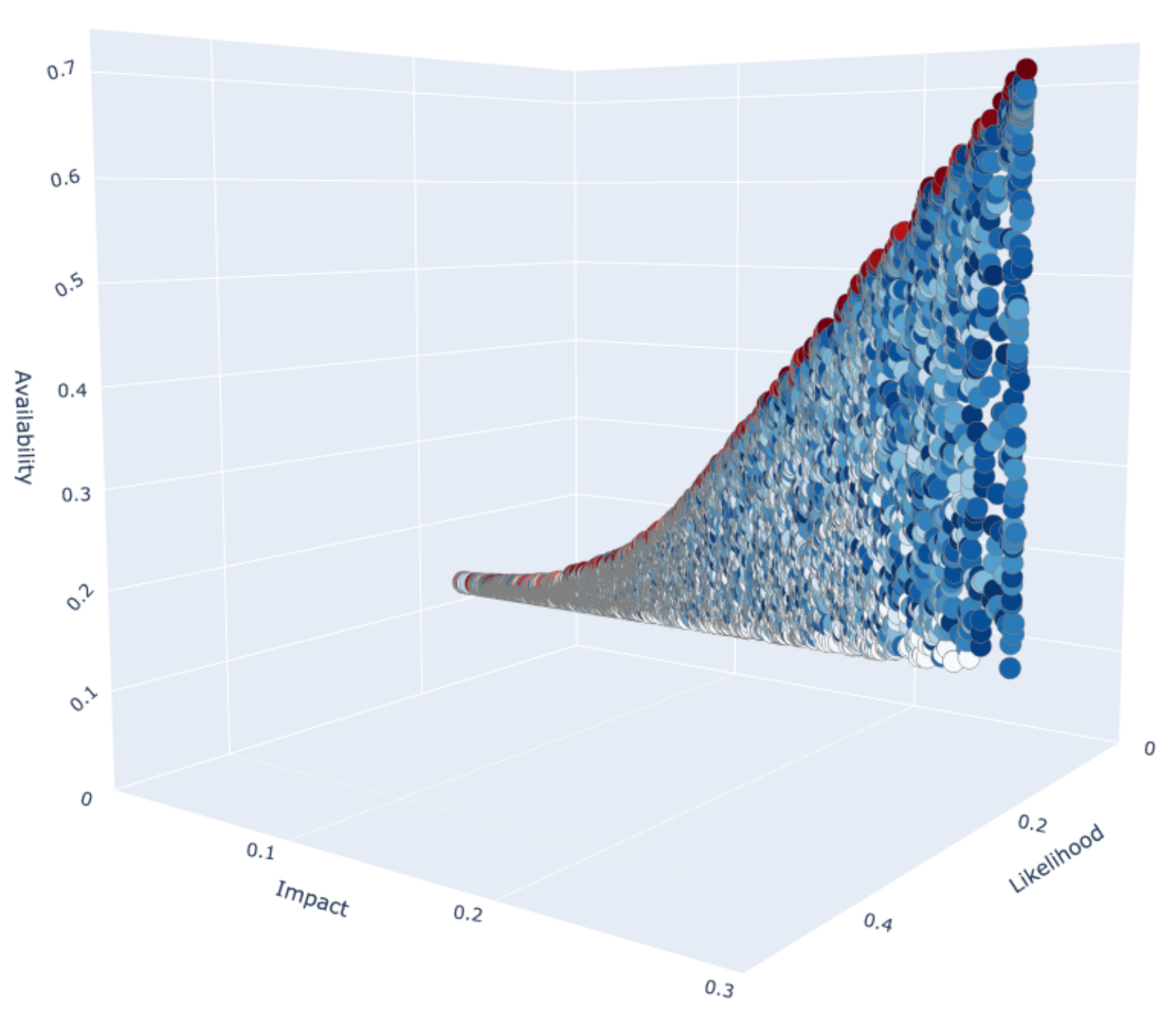}
    \caption{Pareto fronts derived over 10,000 optimisation trials in the BlackEnergy case study}
    \label{fig:case_BlackEnergy_Plot_3D}
\end{figure}

We recommend that practitioners do not interpret this composite risk score in isolation for incident-prioritisation purposes, but to do so alongside complementary metrics instead. A fuller discussion of this interpretive risk, and its implications for deep, multi-stage attack chains more generally, is provided in Section~\ref{sec:discussion}.

\begin{table}[ht]
\centering
\scriptsize
\setlength{\tabcolsep}{2pt}
\renewcommand{\arraystretch}{1.1}
\caption{Top-performing countermeasure portfolio from the BlackEnergy case study}
\begin{tabular}{l l}
\toprule
\textbf{Metric} & \textbf{Value} \\
\midrule
Trial ID & 9683 \\
Countermeasure portfolio & 
V1:0.9901,\ V2:0.9855,\ V3:0.9777, \\
& V4:0.9821,\ V5:0.9954,\ V6:0.9892, \\
& V7:0.9824,\ V8:0.9903,\ V9:0.9713, \\
& V10:0.9722,\ V11:0.9772 \\
Likelihood of successful attack & 0.1486 \\
Probability of severe impact & 0.2899 \\
Availability & 0.6766 \\
\bottomrule
\end{tabular}
\label{tab:case_BlackEnergy_results_3D}
\end{table}

Next, we performed optimisation studies to search for Pareto-optimal countermeasure portfolios to support decision-making in mitigating the BlackEnergy attack. Figure~\ref{fig:case_BlackEnergy_Plot_3D} reveals an increasingly linear relationship between impact probability and system availability at higher impact probability levels.
Additionally, the Pareto fronts appear clustered near a similar value along the attack success likelihood axis, suggesting limited variation in attack success likelihood across trials.

Analysis of the top-performing countermeasure portfolio presented in Table~\ref{tab:case_BlackEnergy_results_3D} highlights a clear trade-off between exposure, impact, and system availability. The model prescribes near maximal intervention across all identified vulnerabilities, with mitigation parameters approaching $1.0$. Despite these intensive efforts, the probability of severe impact remains moderate at $0.2899$. This suggests that in the context of the BlackEnergy attack, aggressive mitigation successfully bolsters system availability but fails to fully suppress the residual risk of severe impact on the CPS -- evidence that structural interdependencies within CPSs can sustain severe impact even when individual node exposure is significantly reduced.

\begin{figure}[ht]
\centering
\begin{tikzpicture}
\begin{axis}[
    width=\linewidth,
    height=4cm,
    ymin=3.5, ymax=9,
    ylabel={Average Rank Position},
    xlabel={Vulnerability Nodes},
    symbolic x coords={V1,V2,V3,V4,V5,V6,V7,V8,V9,V10,V11},
    xtick=data,
    xticklabel style={rotate=45, anchor=east, font=\tiny},
    tick label style={font=\tiny},
    label style={font=\scriptsize},
    grid=both,
    grid style={line width=.1pt, draw=gray!20},
    major grid style={line width=.2pt, draw=gray!40},
    legend style={
        at={(0.5,1.05)},
        anchor=south,
        legend columns=4,
        font=\tiny,
        /tikz/every even column/.append style={column sep=0.2cm}
    },
]

\addplot[color=blue, mark=*, mark size=1.2pt, thick] coordinates {
    (V1,7.23) (V2,6.81) (V3,5.46) (V4,5.18) (V5,5.87) (V6,5.21) 
    (V7,5.31) (V8,5) (V9,6.89) (V10,5.7) (V11,7.34)
};

\addplot[color=red, mark=square*, mark size=1.2pt, thick] coordinates {
    (V1,6.81) (V2,7.51) (V3,5.57) (V4,5.63) (V5,5.33) (V6,5.1) 
    (V7,5.4) (V8,5.2) (V9,6.54) (V10,5.64) (V11,7.27)
};

\addplot[color=green!60!black, mark=triangle*, mark size=1.4pt, thick] coordinates {
    (V1,7.08) (V2,7.39) (V3,5.47) (V4,5.6) (V5,4.52) (V6,5.58) 
    (V7,5.34) (V8,5.3) (V9,6.15) (V10,5.62) (V11,7.95)
};

\addplot[color=orange!90!black, mark=diamond*, mark size=1.4pt, thick] coordinates {
    (V1,6.93) (V2,7.46) (V3,5.04) (V4,6.19) (V5,5.26) (V6,5.19) 
    (V7,4.95) (V8,5.28) (V9,6.56) (V10,6) (V11,7.14)
};

\legend{AF=0.25, AF=0.5, AF=0.75, AF=1.0}

\end{axis}
\end{tikzpicture}
\caption{Average rank positions of mitigation probabilities across 100 runs of 10,000 optimisation trials, repeated across varying AF settings.}
\label{fig:case_BlackEnergy_freq_af}
\end{figure}

Lastly, we performed a frequency-based heuristic analysis of Pareto-optimal portfolios across multiple optimisation runs to identify countermeasure recommendations for mitigating the BlackEnergy attack. As shown in Figure~\ref{fig:case_BlackEnergy_freq_af}, vulnerabilities \texttt{V3}--\texttt{V8} and \texttt{V10} exhibit comparable bar heights, indicating that effective mitigation requires a distributed strategy across multiple vulnerability nodes concurrently. Although this approach demands greater resources, partial implementation yields proportional losses -- implementing only two of these seven top-ranked mitigations (e.g., \texttt{V3} and \texttt{V4}) achieves only $\approx 28.57\%$ ($2/7$) of the full portfolio's effectiveness, leaving critical attack paths unaddressed. Sensitivity analysis further demonstrates that while Pareto-optimal portfolios remain relatively stable, individual nodes show noticeably larger rank variance across varying AF settings compared to the solar PV inverter case study (Figure~\ref{fig:case_SolarPV_freq_af}). This suggests that certain vulnerability nodes in this scenario are more sensitive to attack feasibility, and the recommended mitigation priorities for these nodes can be expected to change across varying AF settings.

In summary, the key findings from the framework's application to the Ukrainian power grid attack are:

\begin{itemize}
    \item Increasingly linear relationship between impact probability and system availability at higher impact probability levels.
    \item Pareto fronts appear clustered near a similar value along the attack success likelihood axis, suggesting limited variation in attack success likelihood across optimisation trials.
    \item Aggressive mitigation improves availability, but does not fully suppress the residual risk of severe impact on the CPS.
    \item Mitigation must adopt a distributed strategy across multiple vulnerability nodes concurrently.
\end{itemize}

\subsection{Railway CBTC System Attack}
\label{cs_cbtc}

\begin{figure}[ht]
    \centering
    \includegraphics[width=0.7\linewidth]{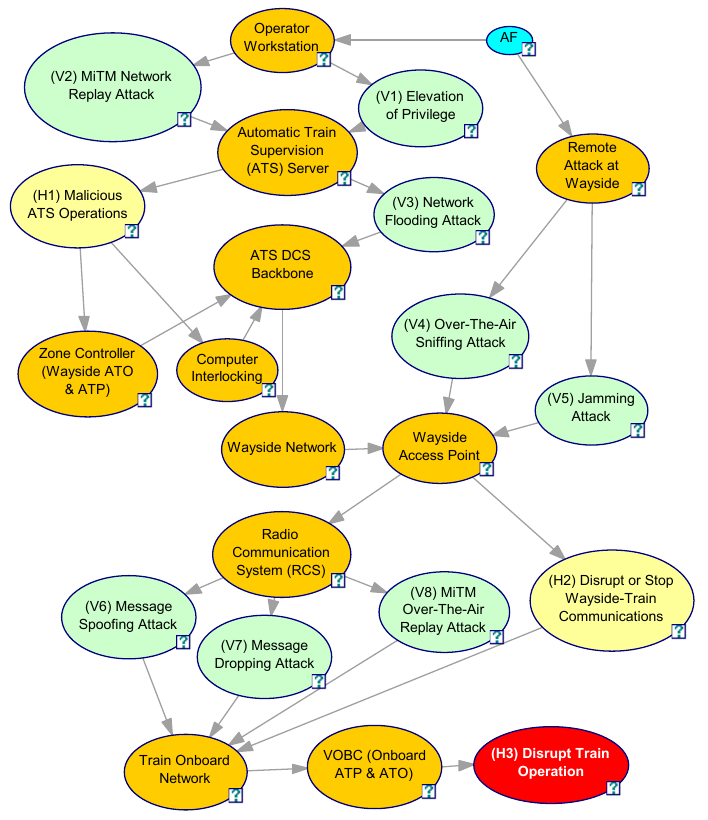}
    \caption{Bayesian Network of a hypothetical cyber-physical attack on a railway CBTC system. A high-resolution version is available in the project's GitHub repository~(\cite{Huang_GitHub}).}
    \label{fig:case_CBTC_BN}
\end{figure}

Next, we shifted our focus to a cyber-physical attack scenario involving a railway CBTC system. To construct a plausible threat model, we drew upon the first author's years of expertise and domain-specific knowledge of railway CBTC systems, complemented by insights from literature that discuss railway-specific risk assessment methods and considerations \cite{Benghename2026}. Synthesising these insights, we formulated a hypothetical cyber-physical attack path that targets CBTC assets through a chain of associated cyber-physical vulnerabilities. These relationships were instantiated within a Bayesian Network model, as illustrated in Figure~\ref{fig:case_CBTC_BN}.

As the entire attack scenario was hypothetical, the vulnerabilities in this case study were not linked to any actual CBTC-related CVE entries. To address this limitation, CVSS 3.1 vectors were constructed by referencing \textit{analogous vulnerabilities} catalogued in the NVD, which bore semantic and structural alignment with the hypothetical vulnerabilities. Initial exposure probability estimates were heuristically derived using Equation~\ref{eq:p_exposure_vuln_noncve}, and subsequently calibrated using the Bayesian latent variable approach outlined in Equation~\ref{eq:posterior_mean_nocve}, taking the assumption of low-confidence estimates.

This approach, however, references CVEs (Table~\ref{tab:cvss_epss_comparison}) that originate from general-purpose ICS, IT, or IoT contexts, and their CVSS vectors may not reflect the exploitability of CBTC-specific equipment given its physical isolation, proprietary protocols, and safety-certified architecture. We treat this cross-domain mapping as a necessary but imperfect substitute in the absence of publicly disclosed CBTC vulnerabilities, and practitioners should interpret the resulting estimates as approximated metrics rather than as validated CBTC risk figures. Validating against domain expert assessments and published railway-specific risk studies is left to future work.

The AF modifier (Equation~\ref{eq:AF_modifier}) was also applied to incorporate attack feasibility in the computation. A value of $\phi = 0.5$ was selected to reflect the contextual conditions typical of an "air-gapped" (i.e., physically isolated) CBTC environment as shown in Figure \ref{fig:case_CBTC_BN}, with the system's security posture assessed as moderate. Cybersecurity practices within this air-gapped CBTC environment were assumed to be moderately controlled but imperfect: removable media used by contractors during routine maintenance were only partially scanned, and internal network segmentation between the safety-critical zone-controller network and the broader maintenance subnet was incomplete, lacking lateral movement monitoring. The adversary is characterised as a competent, moderately resourced insider threat or a persistent team using compromised supply-chain media to execute a multi-stage attack starting at the workstation level. The resulting probability of a successful attack on a vulnerability node is therefore calculated as \( P(A) = P(E) \times 0.5 \).

As the CVSS-based vectors in this scenario were derived entirely from analogous vulnerability references, we performed a comparative analysis against corresponding EPSS-based scores reflecting empirical ``in-the-wild'' attack probabilities. Specifically, we aimed to assess whether the model's output (posterior probabilities) would differ materially between the two scoring systems. For instance, in the case of V1 (CVE-2020-7544), the corresponding CVSS vector (\emph{CVSS:3.1/AV:L/AC:L/PR:L/UI:N/S:U/C:H/I:H/A:H}) yielded $P(E)_{\text{vuln, CVSS}} = 0.55 \times 0.77 \times 0.62 \times 0.85 = 0.2232$, greatly exceeding the corresponding EPSS score of 0.00309.

\begin{table}[htbp]
    \centering
    \scriptsize
    \caption{Comparison of CVSS- and EPSS-based exposure probabilities for each vulnerability}
    \label{tab:cvss_epss_comparison}
    \begin{tabular}{llllc}
        \toprule
        \textbf{ID} & \textbf{CVE} & \textbf{Threat Category} & \textbf{CVSS 3.1} & \textbf{EPSS} \\
        \midrule
        V1 & CVE-2020-7544  & Elevation of Privilege  & 0.2232 & 0.00309 \\
        V2 & CVE-2023-36857 & MiTM Network Replay     & 0.4984 & 0.00309 \\
        V3 & CVE-2024-23814 & Network Flooding        & 0.4984 & 0.00574 \\
        V4 & CVE-2023-0972  & Over-The-Air Sniffing   & 0.4984 & 0.00347 \\
        V5 & CVE-2026-22917 & Jamming                 & 0.4984 & 0.00509 \\
        V6 & CVE-2022-32747 & Message Spoofing        & 0.4984 & 0.00311 \\
        V7 & CVE-2019-10936 & Message Dropping        & 0.4865 & 0.02065 \\
        V8 & CVE-2023-36857 & MiTM Network Replay         & 0.4984 & 0.00309 \\
        \bottomrule
    \end{tabular}
\end{table}

Table~\ref{tab:cvss_epss_comparison} confirms that this pattern holds across all eight vulnerabilities in this scenario, with CVSS-based exposure probabilities consistently exceeding their EPSS counterparts. This suggests that, in this scenario, CVSS over-estimates vulnerability exposure relative to EPSS, which instead reflects the empirical likelihood of near-term exploitation as observed in the wild.

\begin{table}[ht]
    \centering
    \scriptsize
    \caption{Comparison of CVSS- and EPSS-based Joint Conditional Exposure and Impact Probabilities (Railway CBTC)}
    \label{tab:cpt_joint_cbtc_comparison}
    \begin{tabular}{lcc|cc}
    \toprule
    \textbf{State} & \multicolumn{2}{c|}{\textbf{CVSS-Based}} & \multicolumn{2}{c}{\textbf{EPSS-Based}} \\
     & $P(E)$ & $P(I)$ & $P(E)$ & $P(I)$ \\
    \midrule
    \texttt{(0)} & 0.1574 & 1.0000 & 0.1574 & 1.0000 \\
    \texttt{(1)} & 0.8426 & 0.0000 & 0.8426 & 0.0000 \\
    \bottomrule
    \end{tabular}
\end{table}

As shown in Table~\ref{tab:cpt_joint_cbtc_comparison}, the joint conditional probability distributions for exposure ($P(E)$) and impact ($P(I)$) are identical under both the CVSS- and EPSS-based configurations. This indicates that, in this scenario, the differences between CVSS- and EPSS-based exposure probabilities do not translate into materially different downstream risk estimates. This convergence further suggests that the BN's graph structure, rather than the choice of scoring system, is the dominant driver of the posterior outcome. Because the attack path depends on a combination of several upstream conditions, including the individual exposure probabilities of each contributing node, the joint conditional probability is comparatively insensitive to moderate shifts in any single node's exposure probability.

Secondly, the analysis highlights a significant vulnerability within the railway CBTC architecture. While the likelihood of attack success is relatively low ($P(E) = 0.1574$), the propagated impact score reaches its ceiling ($P(I) = 1.0$), reflecting that the attack path terminates directly at the highest-connectivity hazard node in the graph. Consequently, these results suggest that preventive strategies should be prioritised to reduce the probability of entry, as the current system configuration offers no structural buffering once the goal is achieved.

\begin{figure}[ht]
    \centering
    \includegraphics[width=0.75\linewidth]{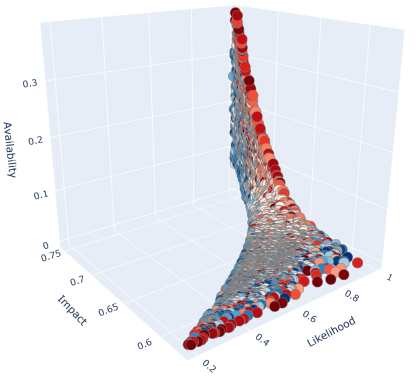}
    \caption{Pareto fronts derived over 10,000 multi-objective optimisation trials in the railway CBTC case study}
    \label{fig:case_CBTC_Plot_3D}
\end{figure}

Next, we conducted optimisation experiments, separately using calibrated CVSS-based priors and EPSS-based priors to evaluate differences in outcomes. In both cases, the derivation of the Pareto fronts (Figure~\ref{fig:case_CBTC_Plot_3D}) illustrates a trade-off between exposure, impact, and availability. The results show a consistent relationship in which the likelihood of severe impact remains elevated across most levels of attack success likelihood. Dense solution clusters are observed near $\textit{Likelihood} \approx 0.2-0.4$, with a sparser tail of solutions extending toward $\textit{Likelihood} \approx 1$. Across these Pareto fronts, impact remains consistently high, while availability is generally low, rising only as attack success likelihood approaches its upper bound. This suggests that the optimisation process prioritises minimising attack success likelihood over reducing impact severity, with availability improving only as a secondary effect at high attack success likelihood.

\begin{table}[ht]
\centering
\scriptsize
\setlength{\tabcolsep}{2pt}
\renewcommand{\arraystretch}{1.2}
\caption{Comparison of top-performing countermeasure portfolios from the railway CBTC case study using calibrated CVSS-based and EPSS-based priors}
\label{tab:CBTC_CVSS_EPSS_comparison}
\begin{tabular}{@{}l l l@{}}
\toprule
\textbf{Metric} & \textbf{CVSS-based (Trial 8361)} & \textbf{EPSS-based (Trial 9234)} \\
\midrule
Countermeasure portfolio & 
V1:0.9986,\ V2:0.9883, &
V1:0.9759,\ V2:0.9778, \\
& V3:0.9788,\ V4:0.9994, & 
V3:0.9792,\ V4:0.9835, \\
\ & V5:0.9929,\ V6:0.9961, &
V5:0.9942,\ V6:0.9873, \\
\ & V7:0.9579,\ V8:0.9914 &
V7:0.9862,\ V8:0.9532 \\
Probability of severe impact & 0.7581 & 0.7559 \\
Availability & 0.4694 & 0.4462 \\
Total execution time & 0h 4m 31s & 0h 4m 23s \\
\bottomrule
\end{tabular}
\end{table}

The top-performing countermeasure portfolios are presented in Table~\ref{tab:CBTC_CVSS_EPSS_comparison}. Both CVSS- and EPSS-based solutions reflect a broad system-wide defence strategy in the scenario of a railway CBTC attack, with near-maximal mitigation across all variables.

Lastly, we analysed Pareto-optimal portfolios obtained from repeated optimisation runs using both calibrated CVSS- and EPSS-based priors, applying a frequency-based ranking to identify which vulnerability nodes were most consistently selected for Pareto-optimal countermeasure portfolios. There were visible differences between results derived from calibrated CVSS- and EPSS-based priors respectively. As illustrated in Figure~\ref{fig:case_CBTC_rank_comparison}, the calibrated CVSS-based distribution exhibits a largely uniform pattern, with \texttt{V3} exhibiting the strongest mitigation effect. In contrast, the EPSS-based distribution is more varied, with \texttt{V2} exhibiting visibly lower mitigation effects relative to other nodes. This marked difference in the CVSS- and EPSS-based distributions may be due to the more heterogeneous empirical attack probabilities that are incorporated in the EPSS-based priors.

\begin{figure}[ht]
\centering
\scriptsize
\begin{subfigure}[b]{0.49\linewidth}
\centering
\begin{tikzpicture}
\begin{axis}[
    ybar, 
    bar width=5pt, 
    width=\textwidth, 
    height=3cm,
    enlarge x limits=0.18, 
    ylabel={Avg. Rank}, 
    xlabel={Nodes},
    symbolic x coords={V1,V2,V3,V4,V5,V6,V7,V8}, 
    xtick=data,
    ymin=0, ymax=6, 
    tick label style={font=\tiny}
]
\addplot+[fill=blue!60] coordinates {
    (V1,5.00) (V2,5.33) (V3,3.47) (V4,4.58)
    (V5,4.61) (V6,4.42) (V7,4.14) (V8,4.45)
};
\end{axis}
\end{tikzpicture}
\caption{CVSS-Based}
\end{subfigure}
\hfill
\begin{subfigure}[b]{0.49\linewidth}
\centering
\begin{tikzpicture}
\begin{axis}[
    ybar, 
    bar width=5pt, 
    width=\textwidth, 
    height=3cm,
    enlarge x limits=0.18, 
    ylabel={Avg. Rank}, 
    xlabel={Nodes},
    symbolic x coords={V1,V2,V3,V4,V5,V6,V7,V8}, 
    xtick=data,
    ymin=0, ymax=6, 
    tick label style={font=\tiny}
]
\addplot+[fill=green!60] coordinates {
    (V1,4.00) (V2,6.80) (V3,2.80) (V4,4.80)
    (V5,4.40) (V6,4.20) (V7,3.80) (V8,5.20)
};
\end{axis}
\end{tikzpicture}
\caption{EPSS-Based}
\end{subfigure}
\caption{Comparison of average mitigation rank positions across 100 optimisation runs. Lower values indicate higher mitigative effectiveness.}
\label{fig:case_CBTC_rank_comparison}
\end{figure}

In summary, the key findings from the framework's application to the railway CBTC system attack are:

\begin{itemize}
    \item No marked differences between conditional probability tables based on CVSS- and EPSS-based priors.
    \item Optimisation results prioritise attack prevention over reducing impact severity, with availability improving only as a secondary effect at high attack success likelihood.
    \item Countermeasure portfolios show near-maximal mitigation across all vulnerabilities, reflecting a system-wide defence approach.
    \item Frequency analysis shows uniform emphasis in countermeasure portfolios derived from calibrated CVSS-based priors, while portfolios derived from EPSS-based priors indicate more targeted mitigation strategies.
\end{itemize}

\subsection{Trade-offs and Practitioner Guidance}

In interpreting the results of our three case studies, scenario-specific behaviours should be distinguished from properties of the framework itself. Patterns such as the shapes of Pareto fronts, the prominence of specific vulnerabilities, or clustering effects in the BlackEnergy and CBTC experiments stem from the particular architectures, threat models, and parameter settings of those scenarios. In contrast, the ability to generate Pareto-optimal countermeasure portfolios, quantify trade-offs between exposure, impact, and availability, and rank mitigation options via frequency-based heuristics are general characteristics of our proposed method that should extend to a broad range of CPS deployments.

These findings provide practical guidance for CPS practitioners regarding the navigation of security trade-offs. The results indicate that complementary vulnerability scoring systems such as CVSS, EPSS, LEV, and KEV can be used flexibly even when vulnerabilities lack CVE identifiers. Specifically, the integration of LEV and KEV scores ensures that confirmed exploitation evidence and local environmental factors take precedence in the prioritisation process. This multidimensional approach, supported by frequency analysis for the comparative evaluation of results, enables more targeted mitigation strategies under operational constraints.

A central insight from our evaluation is that incident mitigation in CPSs inherently involves a three-way trade-off between exposure, impact, and system availability. Across all case studies, the optimisation results reveal regions where reducing the likelihood of successful attack or severe impact begins to incur disproportionately high availability costs, as well as regions where additional mitigation yields diminishing returns in risk reduction. In practice, operators must therefore balance tolerable residual exposure and impact against the need to keep CPS services online, rather than pursuing maximal mitigation in isolation.

\section{Discussion and Future Work}
\label{sec:discussion}

\paragraph{Framework Applicability to Real-World Attacks}

All three case studies, including the solar PV inverter running example in Section~\ref{sec:methodology}, demonstrate the applicability and utility of our proposed framework in supporting adaptive decision-making needs for real-world attacks. As new information becomes available, elements such as CPS assets, probabilistic estimates of attack success likelihood, impact severity, system availability, and candidate countermeasure profiles, comprising mitigation probabilities, are continuously updated via a feedback loop driven by security monitoring and computationally efficient multi-objective optimisation.

From a deployment perspective, the framework presupposes access to architectural knowledge, vulnerability metadata such as CVEs, and attack telemetry to update evidence in the BN graph. It is designed to integrate as an analytical back-end that can be invoked from existing Security Operations Centre (SOC) workflows. In particular, the framework can serve as a complementary decision-support service that consumes alerts and configuration data, rather than replacing existing SOC tooling. Our experiments in this work demonstrate that multi-objective optimisation can be parallelised to ensure runtimes compatible with operational time-frames. However, practical deployments will require engineering trade-offs between model granularity, update frequency, and available compute resources to ensure acceptable overhead in operational settings.

\paragraph{Prior-dominance in Bayesian confidence calibration}

The choice of priors largely determines the behaviour of our proposed Bayesian confidence calibration technique detailed in Section~\ref{subsec:method_exposure_calculation}. From Equation \ref{eq:posterior_mean_nocve}, we observe that the posterior mean $\mu_{\text{post}}$ is a precision-weighted average of the prior mean $\mu_0$ and the CVSS-proxy estimate $P(E)_{\text{vuln, CVSS}}$, represented as follows:

\begin{equation}
\mu_{\text{post}} = w_0 \mu_0 + (1 - w_0)\, P(E)_{\text{vuln, CVSS}}
\label{eq:posterior_mean_weighted}
\end{equation}
\begin{equation}
w_0 = \frac{1/\sigma_0^2}{1/\sigma_0^2 + 1/\sigma^2_{\text{CVSS}}}
\label{eq:posterior_weight}
\end{equation}

where $w_0 \in (0,1)$ is the precision-weighted contribution of the prior, determined by the ratio of the prior variance $\sigma_0^2$ to the CVSS-proxy variance $\sigma^2_{\text{CVSS}}$, not their absolute values.

With a prior $\sigma_0^2 = 0.0025$ against a baseline $\sigma^2_{\text{CVSS}} = 0.04$ (reflecting a neutral Gaussian distribution and a low-confidence CVSS estimate, respectively), the prior carries the bulk of the posterior weight: $w_0 \approx 0.94$. Consequently, $\mu_{\text{post}}$ remains close to $\mu_0$ across all tested configurations (Tables~\ref{tab:sensitivity_analysis_solar} and \ref{tab:sensitivity_analysis_blackenergy}).

The prior's dominance is intentional, not a structural inability to respond to proxy evidence. Because CVSS-proxy estimates for undocumented vulnerabilities are prone to assessor bias \cite{Wunder2024}, we deliberately set a tight prior standard deviation ($\sigma_0 = 0.05$) so that a single noisy CVSS-proxy estimate cannot dominate the posterior when confidence in that estimate is low (e.g., $\sigma_\text{CVSS} = 0.20$). However, this restraint is not absolute, as illustrated in the BlackEnergy scenario (Section~\ref{subsec:cs_blackenergy}). Here, a high-confidence proxy of $P(E)_\text{vuln, CVSS} = 0.306$ that disagrees with the prior ($\mu_0 = 0.5$) still produces a material shift in $\mu_\text{post}$, demonstrating genuine responsiveness to strong evidence.

The $\sigma_0$ value can therefore be regarded as a supporting component for the risk assessment. Setting $\sigma_0$ empirically, from the observed spread of proxy estimates across comparable vulnerabilities, is left as a direction for future work.

\paragraph{Repeatability of Optimisation Outcomes}

An important consideration arising from our research concerns the repeatability of outcomes, such as countermeasure portfolios, produced by multi-objective optimisation trials in response to identical model configurations and threat information. This directly relates to the stability and variance of Pareto fronts across repeated optimisation runs. To evaluate this, we pose the question: \textit{What configuration of trial count is sufficient to ensure that the resulting Pareto front remains relatively consistent across independent multi-objective optimisation runs?}

To address this research question, we evaluated the nearness and clustering behaviour of Pareto fronts generated from 100, 1000, 5000, and 10,000 trials, each repeated over 100 optimisation runs. Comparison of the Pareto fronts across experimental configurations reveals that lower trial counts (e.g. 100) result in considerable dispersion and variability across runs, whereas higher trial counts (e.g. 5000 and 10,000) yield denser, more structurally consistent front distributions across objective space.

Kernel Density Estimation (KDE)~\cite{Chen2017} was employed, following unimodality validation across attack success likelihood, impact probability, and availability, to quantify and compare cluster densities (Table~\ref{tab:discussion-kde-density-comparison}). The results confirm that 5,000 trials suffice to produce consistent, structurally reliable Pareto fronts across independent optimisation runs. However, applications requiring finer resolution or density differentiation (e.g., countermeasure ranking or adaptive planning) may benefit from a 10,000-trial configuration, albeit with higher execution time.

\begin{table}[t]
\centering
\scriptsize
\caption{Comparison of KDE-derived density metrics for Pareto-optimal solutions across trial configurations (100 runs each)}
\begin{tabular}{l r r r r}
\toprule
\textbf{Metric} & \textbf{100 trials} & \textbf{1,000 trials} & \textbf{5,000 trials} & \textbf{10,000 trials} \\
\midrule
Average KDE              & 601.17   & 543.70   & 359.65   & 395.99 \\
Minimum density          & 136.64   & 51.13    & 31.01    & 45.91 \\
Maximum density          & 1130.66  & 1041.85  & 698.14   & 887.52 \\
Density variance         & 62{,}435.93 & 84{,}427.79 & 36{,}824.83 & 56{,}210.58 \\
Density entropy          & 4.51     & 4.44     & 4.43     & 4.40 \\
\bottomrule
\end{tabular}
\label{tab:discussion-kde-density-comparison}
\end{table}

\paragraph{Concurrency and Computational Efficiency}

The optimisation experiments conducted in this research underscore the critical role of concurrency in enhancing computational efficiency and responsiveness for multi-objective decision-making in CPSs. Using the Python concurrent.futures module, we achieved substantial reductions in cumulative optimisation time (Table~\ref{tab:discussion-computation-comparison}).

\begin{table}[ht]
\centering
\caption{Comparison of execution times for 3 runs of 10,000 trials each}
\begin{tabular}{l c c}
\toprule
\textbf{Configuration} & \textbf{Optuna} & \textbf{Optuna + concurrent.futures} \\
\midrule
Total execution time & 15 mins 3 secs & 5 mins 10 secs \\
\bottomrule
\end{tabular}
\label{tab:discussion-computation-comparison}
\end{table}

\paragraph{Threats To Validity}

A limitation of this evaluation is the absence of ground-truth benchmarks: none of the three case studies are scored against an independently verified "correct" countermeasure portfolio. This is difficult to avoid in practice, since documented remediation sequences for real incidents are rarely published in sufficient detail, red-team outcomes are typically held privately, and the CBTC scenario is hypothetical by construction. We therefore present the three case studies as evidence that the framework produces Pareto-optimal, operationally-interpretable portfolios across diverse CPS domains, not as evidence that those portfolios are optimal in an absolute sense. Establishing the latter would require proprietary incident response data or a dedicated red-team evaluation, which we leave to subsequent work.

\paragraph{Practical Implications}

It is important to distinguish our proposed framework from existing Operational Technology (OT) security platforms such as Claroty \cite{Claroty2026}, Dragos \cite{Dragos2024}, and others. Whereas these platforms are designed for incident detection and response, our proposed framework is more appropriately positioned as an \textit{adaptive decision-support tool} for CPS practitioners, particularly in situations where evidence is incomplete and system states are evolving. In this respect, our proposed framework complements, rather than replaces, existing OT security platforms.

Secondly, we emphasise that the three case studies in this work are intended to demonstrate the applicability and mechanics of our proposed framework, rather than to provide a comparative or ground-truth evaluation against alternative decision-support approaches. The BlackEnergy and solar PV inverter scenarios are grounded in real, CVE-linked vulnerabilities and publicly documented incident reports, but the railway CBTC scenario is, by necessity, fully hypothetical, as no equivalent public incident or CVE record exists for this CPS. The conjunctive propagation of posterior probabilities across BlackEnergy's multi-stage attack chain yields a low composite risk score despite being the only scenario grounded in a confirmed real-world blackout. This reflects the framework's modelling of attack success likelihood under a specified security posture rather than historical impact severity, and practitioners should be aware that deep, multi-stage attack paths may systematically produce lower marginal exposure probabilities than shallower, opportunistic attack trees.

We clarify that this does not indicate the framework is miscalibrated: recomputing the risk score $R$ under a weaker security posture (e.g., $\phi$ approaching 1.0) yields proportionally higher scores, confirming the model responds correctly to its inputs. Rather, the result highlights a limitation of using a single composite score for incident prioritisation in deep attack chains. We recommend that practitioners interpret scores with $P(I)$ as a complementary metric: in the BlackEnergy scenario, $P(I) = 0.2393$ (Table~\ref{tab:cpt_joint_blackenergy}) signals that a realised attack path carries considerable consequence, information the low composite risk score alone obscures.

\paragraph{Future Work}

To further strengthen decision-making capabilities in CPS environments, future work should explore several complementary directions that address the key challenges identified in our research.

First, our research highlighted the promise of frequency-based heuristics as a supplementary analytical layer beyond individual optimisation outcomes. However, their empirical nature prompts a critical question: \textit{To what extent can such heuristically derived recommendations be relied upon in safety-critical CPS environments?} Addressing this question presents a valuable avenue for future work, particularly through comparative evaluation with empirical data. Future work could also extend optimisation to explicitly account for resilience alongside system availability.

Second, the stability of Pareto fronts across repeated optimisation trials remains central to ensuring decision robustness. Low trial counts were shown to produce high variance in outcomes, which may undermine the trustworthiness of mitigation recommendations. Future investigations should therefore develop convergence diagnostics to establish sufficiency thresholds for trial configurations, and to evaluate optimisation solutions.

Third, each AF modifier value in this paper was elicited from a single analyst per scenario, so we do not yet know how much it would vary across analysts assessing the same evidence. The natural next step is to have two or three analysts independently score AF for the same case study and measure agreement directly, then check whether disagreement at that level changes which countermeasures the framework recommends.

Fourth, the incorporation of hybrid confidence-calibrated exposure probability assessments presents a promising direction for performing vulnerability assessment in practical deployments. Such hybrid approaches are particularly relevant in guiding real-world incident mitigation strategies that must adapt in real time to evolving attacks and dynamic threat intelligence. Furthermore, the approach is inherently extensible, facilitating the integration of diverse vulnerability scoring systems such as CVSS, EPSS, LEV, and KEV to refine the quantification of exposure estimates. Future work could further leverage this multidimensional approach to refine decision support, enabling a more comprehensive and data-informed basis for prioritising countermeasures as new threat intelligence becomes available.

Finally, architectural improvements and mechanisms could be explored to dynamically rebalance the factors employed in impact calculations (as outlined in Section~\ref{subsec:method_impact_calculation}), thereby enabling the generation of countermeasure portfolios that reflect evolving runtime priorities. This may include structural enhancements such as the incorporation of noisy-OR gates to improve agility in assimilating new evidence, and to better support adaptive decision-making.

Collectively, these directions aim to further improve an adaptive decision support framework for mitigating cyber incidents in CPSs.

\begin{table*}[t]
\caption{Comparison of optimisation frameworks for CPS incident response}
\label{tab:optimisation_framework_comparison}
\centering
\footnotesize
\renewcommand{\arraystretch}{1.15}
\begin{tabularx}{\textwidth}{@{}l X X X X X@{}}
\hline
\textbf{Framework} & \textbf{Vulnerability/Threat Input} & \textbf{Uncertainty Handling} & \textbf{Optimisation Objective(s)} & \textbf{Adaptability Mechanism} & \textbf{Evaluation} \\
\hline
Li et al.~\cite{Li2018} & CVSS-derived exploitability, fixed & None -- point probabilities treated as ground truth & Single-objective intrusion-response ranking & Reactive to detected intrusion; exposure not revised as evidence accrues & Simulated ICS testbed \\
Żebrowski et al.~\cite{Zebrowski2022} & Expert-elicited probabilities & None -- static, fixed at model-build time & Pareto-efficient portfolio under fixed budget & Offline BN; no update mechanism & Single case study; no ground truth \\
Zaman et al.~\cite{Zaman2022} & Disruption-event states (not vulnerability-scoring) & Not explicit & MDP-based response-action selection & Adaptive within MDP state space; smart-building domain, not vulnerability-driven & Smart-building case study \\
Shimizu \& Hashimoto~\cite{Shimizu2025} & CVSS + EPSS + KEV, deterministically chained & None -- scores combined without reliability calibration & Single-objective vulnerability prioritisation & Static; targets enterprise IT, not CPS constraints & Not evaluated on CPS \\
\textbf{Proposed} & CVSS + EPSS + LEV + KEV, Bayesian-calibrated & Epistemic uncertainty surfaced as posterior variance metric & Joint attack likelihood, impact, availability (multi-objective) & BN attributes updated from evidence; frequency heuristics re-prioritise across runs & Three CPS case studies (real + hypothetical) \\
\hline
\end{tabularx}
\end{table*}

\section{Related Work}
\label{sec:related_work}

\paragraph{Probabilistic Modelling in Cybersecurity}

BN graphs are highly effective in synthesising heterogeneous data for cybersecurity modelling, particularly when historical attack data is sparse~\cite{Chockalingam2017,Huang2025_SLR}. Prior studies~\cite{Xie2010,Poolsappasit2011,Wang2017,Khosravi2020} underscore their utility in dynamic decision-making. Bhosale et al.~\cite{Bhosale2023,Bhosale2024} further integrated hazard, vulnerability, and asset connectivity into Bayesian Belief Networks (BBNs) using AutomationML for data exchange. However, existing frameworks often overlook the operational trade-offs of security mitigations, such as asset degradation or failure from software patching, and do not account for legacy system constraints and operational limitations. Furthermore, a heavy reliance on CVSS v3.1 scores limits adaptability to shifting threat landscapes where probabilistic insights from EPSS~\cite{EPSS2021} are required. Our research addresses these gaps by proposing a consequence-aware, iterative framework that integrates confidence-calibrated exposure estimation within an adaptive decision-support pipeline combining BN-based risk propagation, multi-objective optimisation, and frequency-based heuristic prioritisation.

\paragraph{Consequence-Driven Security Assessment}

Beyond modelling attack propagation, effective CPS decision-making must account for the consequences of cyber-\\physical incidents. Kim et al.~\cite{Kim2022} incorporate a consequence layer within BN graphs to quantify impact propagation in exploited ICSs, detailing functional interdependencies and infrastructure-wide effects. In parallel, Amro et al.~\cite{Amro2023} propose a consequence-driven framework rooted in Failure Mode, Effects, and Criticality Analysis (FMECA) that aligns well with CPS environments. Although these approaches support iterative assessment, their transformation into adaptive tools hinges on modelling realistic, multi-stage attacks. In contrast to these approaches, our research integrates system-level functional dependencies into BN graphs informed by empirical failure modes, explicitly modelling uncertainty within exposure computations to enable responsive decision-making.

\paragraph{Decision Support for CPS Incidents}

Research has explored diverse methodologies for CPS decision-making under uncertainty, including Zaman et al.'s~\cite{Zaman2022} MDP (Markov Decision Process) framework for hazard response and Javornik et al.'s~\cite{Javornik2022} mission-centric DSS for resilient IT configuration. Both provide structured resilience assessments but lack specific focus on the adversarial tactics and non-linear cascading risks inherent in CPS incidents -- a gap our BN-based approach addresses by modelling risk propagation that reflects evolving threat scenarios.

\paragraph{Multi-Objective Optimisation}

Decision-making in CPS incident response must balance cybersecurity, safety, and operational resilience. Rehak et al. \cite{Rehak2019} introduce a methodology for assessing critical infrastructure resilience by evaluating robustness, recovery, and adaptive capacity. However, their model is essentially static and focused on pre-incident conditions. Li et al.~\cite{Li2018} propose a multi-objective optimisation framework for intrusion response in ICSs, prioritising Pareto-optimal strategies, yet it does not account for unknown vulnerabilities or emerging vectors. Similarly, {\.Z}ebrowski et al.~\cite{Zebrowski2022} use BN graphs to assess cascading impacts but assume static configurations. While Shimizu et al.~\cite{Shimizu2025} proposed a vulnerability management chaining framework integrating CVSS, EPSS, and KEV metrics, their work is tailored to enterprise IT and lacks the multi-objective weighting necessary to balance the high-availability requirements of CPS environments. Our research addresses these limitations by using a Bayesian confidence calibration technique to model emerging threat uncertainty and implementing an iterative multi-objective optimisation cycle to tailor strategies for CPS incident mitigation. Table~\ref{tab:optimisation_framework_comparison} summarises how our proposed framework relates to comparable prior approaches discussed above.

\paragraph{Railway CBTC Systems}

Several studies have examined the security of railway signalling, with varying emphasis on CBTC technologies. Xu et al.~\cite{Xu2015} present a simulation platform grounded in real-world CBTC characteristics, while Yu et al.~\cite{Yu2023} offer a comprehensive review of the architecture, cybersecurity threats, and protection strategies specific to these systems. In contrast, other research focuses on broader or legacy signalling components. For instance, Schlehuber et al.~\cite{Schlehuber2017} examine interlocking systems without addressing CBTC explicitly, and Unger et al.~\cite{Unger2023} survey railway attack scenarios and countermeasure maturity across the sector more generally. Regarding systemic vulnerabilities, research has highlighted the risks inherent in the transition to wireless-based control. Farooq et al.~\cite{Farooq2017} analyse over-the-air communications in CBTC, identifying interference susceptibility in Wi-Fi-based implementations. Oliveira et al.~\cite{Oliveira2020} expand on this by using Failure Mode and Effects Analysis (FMEA) to uncover risks such as message spoofing and Man-in-the-Middle (MitM) attacks that compromise communication integrity. Similarly, Soderi et al.~\cite{Soderi2023} outline wireless attack scenarios with demonstrated impacts on railway safety. These findings directly inform the adversarial modelling applied in our CBTC case study, ensuring the evaluation reflects the unique cyber-physical risks inherent in modern automated rail operations.

\section{Conclusion}
\label{sec:conclusion}

This paper presented an adaptive decision support framework to help practitioners mitigate cyber-physical incidents in CPS environments -- a critical need as CPS integration into essential infrastructure grows alongside a diversifying threat landscape. A key contribution is using BN graphs, enriched with system and vulnerability metadata, to represent complex attack scenarios. Encoded in a DSL, these models enable dynamic updates that support adaptive decision-making as incidents unfold. The framework also employs a hybrid exposure probability estimation technique to handle vulnerability uncertainty, especially in legacy CPS assets with incomplete data. This method integrates complementary vulnerability scores via Bayesian confidence calibration, bounding proxy-score bias to produce uncertainty-aware reporting metrics for supporting practitioners with downstream risk assessment. Incorporating uncertainty inputs directly into the decision-support pipeline remains future work.

The framework introduces a decision-support methodology for countermeasure selection based on multi-objective optimisation and frequency-based heuristic analysis. By evaluating mitigation probabilities within a structured search space, the system generates Pareto-optimal portfolios balancing security, safety, and operational resilience. Frequency-based heuristics across repeated optimisation runs translate complex model outputs into actionable, constraint-aware operational guidance.

Application across three representative scenarios: the 2015 Ukrainian power grid (BlackEnergy) attack, a solar PV inverter system, and a railway signalling network, demonstrates the framework's versatility against multi-path and multi-agent CPS threats. To ensure transparency, reproducibility, and further extension, all models, source code, and reference datasets are publicly available~\cite{Huang_GitHub}.

\bibliographystyle{IEEEtran}
\bibliography{references}

\begin{IEEEbiography}[{\includegraphics[width=1in,height=1.25in,clip,keepaspectratio]{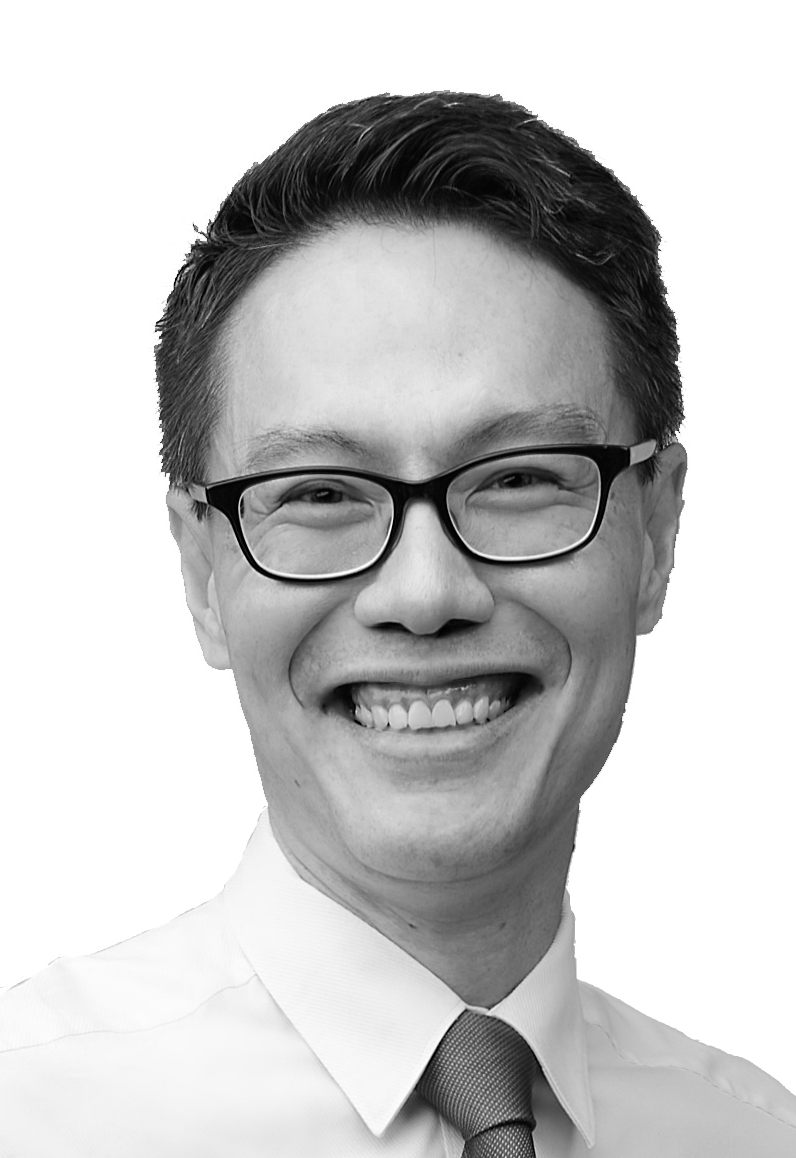}}]{SHAOFEI HUANG} is the Group Chief Information Security Officer at SMRT Corporation. He earned his Master's degree in Business Administration from the University of Leicester in 2002 and Bachelor's degree in Engineering from the University of Tokyo in 1998. Currently, he is pursuing his Doctor of Engineering degree with the School of Computing and Information Systems at Singapore Management University (SMU). His research focuses on security modelling, anomaly detection, and the self-healing capabilities of cyber-physical systems.
\end{IEEEbiography}

\begin{IEEEbiography}[{\includegraphics[width=1in,height=1.25in,clip,keepaspectratio]{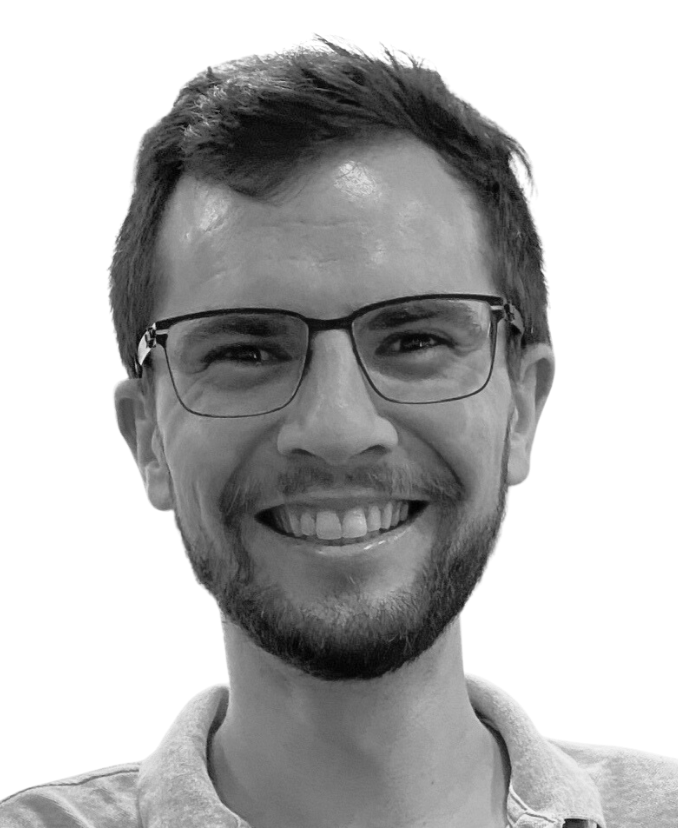}}]{CHRISTOPHER M. POSKITT} is an Associate Professor of Computer Science (Education) at Singapore Management University (SMU), where he is part of the Centre for Research on Intelligent Software Engineering. Prior to SMU, he held postdoctoral research positions at ETH Z\"{u}rich and SUTD, and obtained his PhD in Computer Science from the University of York (2014). His research broadly addresses the problem of engineering correct and secure software, especially in the context of cyber-physical systems (e.g.~industrial control systems, autonomous vehicles). In addition to software engineering, his research interests span formal methods, cybersecurity, and computer science education.
\end{IEEEbiography}

\begin{IEEEbiography}[{\includegraphics[width=1in,height=1.25in,clip,keepaspectratio]{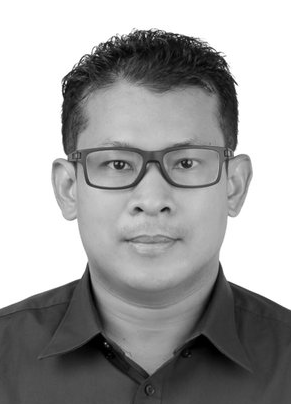}}]{LWIN KHIN SHAR} (Member, IEEE) received the Ph.D. degree in software engineering from Nanyang Technological University (NTU), Singapore, in 2014. He is an Associate Professor with the School of Computing and Information Systems, Singapore Management University, Singapore. He was a Postdoctoral Research Associate with SnT of the University of Luxembourg, Esch-sur-Alzette, Luxembourg, and then a Research Scientist with NTU. His research interests include software engineering, security \& privacy, and machine learning, while specializing in analysis of web \& mobile applications and recently cyber-physical systems for detecting security vulnerabilities, privacy issues, malware, and anomalies. He is author or co-author of more than 40 research papers published in international journals and conferences/workshops, including top venues (e.g., IEEE Transactions on Software Engineering, IEEE Transactions on Dependable and Secure Computing, Empirical Software Engineering, International Conference on Software Engineering, Foundations of Software Engineering, Automated Software Engineering). In his research, he often collaborates with industry partners spanning from healthcare and traffic management to Government sectors.
\end{IEEEbiography}

\EOD

\end{document}